\documentclass[aps,pra,amsmath,showkeys,amssymb,preprint,revsymb]{revtex4-1}
\usepackage{graphicx}
\usepackage{bm}
\usepackage[utf8]{inputenc}
\usepackage[T1]{fontenc}
\usepackage{mathptmx}
\DeclareMathAlphabet{\mathcal}{OMS}{cmsy}{m}{n}
\DeclareMathAlphabet\mathbfcal{OMS}{cmsy}{b}{n}
\usepackage{etoolbox}
 
\usepackage{xcolor}
\usepackage{epstopdf}
\usepackage{caption}
\usepackage{float}
\usepackage{subcaption}
\usepackage{hyperref}
\hypersetup{
    colorlinks,
    citecolor=blue,
    filecolor=blue,
    linkcolor=blue,
    urlcolor=blue}
\begin{document}

\title{Dynamical generation of macroscale magnetic fields and fast flows in a four-component astrophysical plasma}
\author{Usman Shazad}
\email{usmangondle@gmail.com}
\author{M. Iqbal}
\affiliation{Department of Physics, University of Engineering and Technology, Lahore
54890, Pakistan}
\begin{abstract}
We explore the possibility of the generation or amplification of macroscale magnetic fields and flows in a four-component astrophysical dusty plasma composed of mobile massless electrons and positrons, inertial positive ions and negatively charged static dust particles. The investigation demonstrates that when microscopic turbulent ambient plasma energy is predominantly kinetic, a straight dynamo (DY) mechanism is feasible. Conversely, a unified reverse-dynamo/dynamo (RDY/DY) mechanism is possible when the microscopic turbulent ambient plasma energy is primarily magnetic. Additionally, the evolution of Alfv\'{e}n Mach numbers at the macro- and microscale are significantly affected by plasma species densities and invariant helicities. The potential implications of the present study for astrophysical settings are also highlighted.
\end{abstract}

\maketitle
\section{Introduction} \label{s1}
The generation of large-scale magnetic fields from predominantly small-scale velocity fields characterizes the conventional dynamo (DY) mechanism. It involves the emergence of large-scale magnetic fields from initially turbulent systems \cite{Moffat1978}. The DY action appears to be a widely observed phenomenon, occurring in both fusion devices and astrophysics. For instance, the magnetic self-organization process observed in the reverse field pinches is a clear example of the DY in operation \cite{Bodin1990,Chahine2022}. Exploring the potential for interactions that could lead to effective DY action is a highly active area of research in the realm of plasma astrophysics. Understanding the complex plasma processes occurring in astrophysical settings would be a challenge without knowledge of the magnetic field structures \cite{Zweibel1997,Brandenburg2005,Kulsrud2008,Durrer2013,Roberts2013,Helander2016,Rincon2019}. Standard DY theories focus on the creation of large-scale magnetic fields in electrically charged fluids. Theoretical frameworks regarding DY mechanisms have gradually incorporated more sophisticated physics models, advancing from the kinematic to the magnetohydrodynamics (MHD) and, more recently, the Hall MHD (HMHD) dynamo mechanisms. In the more recent theories, the velocity field cannot be forced from outside, as it is in the kinematic case, but instead develops through interaction with the magnetic field. It is natural for both MHD and HMHD dynamo theories to consider the simultaneous evolution of the magnetic and velocity fields.

If the turbulence on a microscale has the potential to create magnetic fields on a macroscale, then it is possible for the turbulence to also generate plasma flows on a macroscale \cite{Mininni2002,Mininni2003,Mininni2003a}. In addition, the turbulent amplification also destroys or dissipates the structures or magnetic elements before they can fully evolve, resulting in substantial flows or causing heating \cite{Bellot2001,Blackman2004,Socas2005}. \citeauthor{Mahajan2005} (2005) proposed the idea of a reverse-dynamo (RDY), which results in the generation of large-scale flows that are simultaneously driven by microscopic fields and flows. Thus, if the process of converting microscale kinetic energy into large-scale magnetic energy is referred to as a "DY", then the inverse process of converting microscale magnetic energy into large-scale kinetic energy could be labeled a "RDY". So extending the definitions is advantageous, as the DY (RDY) process entails the generation of a magnetic field (flow) on a large scale, irrespective of the interplay between microscale magnetic and kinetic energy. Importantly, the investigation also highlighted the simultaneous operation of DY and RDY processes in a simple HMHD system, generating a large-scale magnetic field and plasma flow.
The strength of the macroscopic flow, whether it is weak (sub-Alfv\'{e}nic) or strong (super-Alfv\'{e}nic) in relation to the macroscopic field, was determined by the composition of the turbulent energy. The study also examined the conditions under which one process predominates over the other and deduced the relationships between the generated fields and the flows \cite{Mahajan2005}. Subsequently, \citeauthor{Lingam2015} (2015) proposed that the HMHD-based theory can be characterized as a unified DY/RDY mechanism, in which magnetic fields and flows simultaneously emanate from a given kinetic or magnetic source of short-scale energy. The unified DY/RDY mechanism derived from HMHD also incorporates an inherent length scale known as the ion skin depth, which facilitates the accurate normalization and classification of both microscopic and macroscopic scales. Moreover, it is also hypothesized that numerous observed astrophysical outflows with very large Alfv\'{e}n Mach numbers originate from an efficient RDY \cite{Lingam2015}.
Importantly, an investigation by \citeauthor{Brandenburg2019} (2019) has also revealed a novel characteristic of DYs operating at high magnetic Prandtl numbers: the reversal of magnetic energy into kinetic energy at small scales or high wave numbers. That is indicative of a RDY mechanism. Instead of dissipating energy through very small current sheets, it is responsible for the dissipation of energy by viscous heating \cite{Brandenburg2019}.

In a recent work, Kotorashvili \textit{et al}. (2020) investigated a dense degenerate two-fluid (electron-ion and electron-positron) plasma system in order to obtain the analytical relations governing the unified DY/RDy mechanism. This mechanism involves the generation or amplification of fast macroscale plasma flows in astrophysical systems that initially possess turbulent magnetic or velocity fields at the microscale \cite{Kotorashvili2020}. Similarly, in another study by \citeauthor{Kotorashvili2022} (2022), the unified DY/RDY mechanism and its implications were investigated in a three-component relativistic degenerate two-temperature electron-ion plasma. The plasma being studied was composed of a bulk degenerate electron-ion fluid with a small fraction of classical relativistic hot electrons. These plasmas might be present in white dwarfs accreting hot astrophysical flow or in binary systems. The study also examined the impact of relativistic degeneracy, relativistic temperature of classical hot electrons and density of hot electrons on the unified DY/RDY process \cite{Kotorashvili2022}.

The objective of the present study is to explore the unified DY/RDY mechanism in electron-positron-ion dusty (EPID) plasma. There are numerous electron-ion plasma systems where positron or dust species, or both, can be present as a result of various mechanisms, such as pair production\cite{Murphy2005}, thermal heating, radiative heating, etc \cite{Shukla2002}. EPID plasmas have been observed in various astrophysical environments, such as the galactic center \cite{Zurek1985}, active galactic nuclei (AGN) \cite{Rees1984,Lightman1987,Miller1988,Blandford2019,Czerny2023}, pulsar magnetosphere \cite{Sturrock1971,Ruderman1975,Michel1982}, supernova environments\cite{Tajima,Alfven1981,Shukla2004}, interstellar medium \cite{Shukla2008,Higdon2009}, the earth's magnetosphere\cite{Gusev2001} and solar atmosphere \cite{Murphy2005,Horanyi1996}, as well as laboratory experiments \cite{Surko1990,Krasheninnikov2005,Guanying2017}. It is also important to highlight that, in last few years, there has been a significant amount of research dedicated to studying nonlinear structures and wave propagation in EPID plasmas \cite{Banerjee2016,Paul2017,Singh2017,El-kalaawy2018,Dev2020,Haque2020,Rahman2021,Halder2023}.

In the present work, by considering a four-component magnetized EPID plasma that consists of massless electron and positron species, inertial positive ions and stationary negatively charged dust particles, we formulate an equation and dispersion relation for a unified DY/RDY mechanism. Additionally, the impact of densities of dust and positron species in these processes has been explored, and findings show that densities of plasma species have a significant impact on the generation of large-scale fields and flows in addition to the helicities of the plasma system. In terms of the originality of the present work, as far as we know, no one has studied the unified DY/RDY mechanisms in a four-component EPID plasma. Also, the derived equations governing the unified DY/RDY processes differ from prior studies pertaining to two- and three-component plasmas \cite{Mahajan2005,Lingam2015,Kotorashvili2020,Kotorashvili2022}. 

This paper is arranged in the following manner: In Sec. \ref{s2}, from the model
equations for the EPID plasma, an equation for the unified DY/RDY mechanism
as well as the dispersion relation is derived. The numerical analysis of the
unified DY/RDY mechanism by considering some arbitrary values of plasma
parameters is presented in Sec. \ref{s3}. In the final section, a summary of the
current investigations is provided.

\section{Theoretical model} \label{s2}

We consider an incompressible, quasi-neutral and magnetized multispecies dusty
plasma whose constituents are mobile electrons ($e$), positrons ($p$),
singly ionized positive ions ($i$) and negatively charged stationary dust
particles ($d$). The equations of motion for dynamic $\alpha -$plasma
species ($\alpha =e$, $p$, $i$) can be stated as
\begin{equation}
\frac{\partial \mathbf{v}_{\alpha }}{\partial t}+\left( \mathbf{v}_{\alpha
}\cdot \mathbf{\nabla }\right) \mathbf{v}_{\alpha }=\frac{q_{\alpha }}{
m_{\alpha }}\left( \mathbf{E}+\frac{\mathbf{v}_{\alpha }\times \mathbf{b}}{c}
\right) -\frac{1}{m_{\alpha }n_{\alpha }}\mathbf{\nabla }p_{\alpha },
\label{e1}
\end{equation}
where $m_{\alpha }$, $n_{\alpha }$, $\mathbf{v}_{\alpha }$, $q_{\alpha }$, $p_{\alpha }=n_{\alpha }T_{\alpha }$ and $T_{\alpha }$ denotes mass, number density, flow velocity, electric
charge, thermal pressure and temperature, respectively. Also, $c$, $\mathbf{E}$ and $
\mathbf{b}$ represent the speed of light in vacuum, as well as electric and
magnetic fields, respectively. Since $\left( \mathbf{v}_{\alpha }\cdot 
\mathbf{\nabla }\right) \mathbf{v}_{\alpha }=\mathbf{\nabla }\left(
v_{\alpha }^{2}/2\right) -\mathbf{v}_{\alpha }\times \left( \mathbf{\nabla
\times v}_{\alpha }\right) $, $\mathbf{E}=\mathbf{-\nabla }\varphi
-c^{-1}\left( \partial \mathbf{A}/\partial t\right) $ and $\mathbf{\nabla
\times A=b}$ (where $\varphi$ is electric potential and $\mathbf{A}$ is magnetic vector potential), then the Eq. (\ref{e1}) can be expressed as
\begin{equation}
\frac{\partial }{\partial t}\left( \mathbf{v}_{\alpha }+\frac{q_{\alpha }}{
m_{\alpha }c}\mathbf{A}\right) =\mathbf{v}_{\alpha }\times \left( \mathbf{
\nabla \times v}_{\alpha }+\frac{q_{\alpha }}{m_{\alpha }c}\mathbf{b}\right)
-\mathbf{\nabla }\Psi_{\alpha},
\label{e2}
\end{equation}
where $\Psi_{\alpha}=(v_{\alpha }^{2}/2)+(q_{\alpha}\varphi/m_{\alpha})+(p_{\alpha}/m_{\alpha}n_{\alpha})$. To express the equations of motion (\ref{e2}) in dimensionless form, we
employ the following set of dimensionless variables: $\mathbf{x}=l_{i}%
\widehat{\mathbf{x}}$, $\mathbf{b}=B_{0}\widehat{\mathbf{b}}$, $\mathbf{v}%
_{\alpha }=v_{A}\widehat{\mathbf{v}}_{\alpha }$, $t=\left(
l_{i}/v_{A}\right) \widehat{t}$, $\mathbf{A}=\left( m_{i}cv_{A}/e\right) 
\widehat{\mathbf{A}}$, $\varphi =\left( B_{0}^{2}/4\pi n_{i}e\right) 
\widehat{\varphi }$, $p_{\alpha }=\left( B_{0}^{2}/4\pi \right) \widehat{p}%
_{\alpha }$, $l_{i}=\sqrt{m_{i}c^{2}/4\pi n_{i}e^{2}}$ and $v_{A}=B_{0}/%
\sqrt{4\pi m_{i}n_{i}}$; where $B_{0}$, $l_{i}$, $v_{A}$, $e$, $m_{i}$ and $%
n_{i}$ represent some arbitrary value of ambient magnetic field, ion skin
depth, Alfv\'{e}n velocity, elementary charge, mass and density of ion species,
respectively. It is worth noting that in the current study, we will be
focusing on electron and positron species with negligible mass ($m_{e,p}\ll
m_{i}$). Therefore, the equations of motion for electron, positron and ion
species in dimensionless form can be given as
\begin{equation}
\frac{\partial \mathbf{A}}{\partial t}=\mathbf{v}_{e}\times \mathbf{b}-
\mathbf{\nabla }\left( \varphi -\frac{p_{e}}{N_{e}}\right),  \label{n1}
\end{equation}
\begin{equation}
\frac{\partial \mathbf{A}}{\partial t}=\mathbf{v}_{p}\times \mathbf{b}-
\mathbf{\nabla }\left( \varphi +\frac{p_{p}}{N_{p}}\right) , \label{n2}
\end{equation}
\begin{equation}
\frac{\partial }{\partial t}\left( \mathbf{v}+\mathbf{A}\right) =\mathbf{v}
\times \left( \mathbf{\nabla \times v}+\mathbf{b}\right) -\mathbf{\nabla }
\left( \frac{1}{2}v^{2}+\varphi +p_{i}\right) ,  \label{n3}
\end{equation}
where $N_{e}=n_{e}/n_{i}$, $N_{p}=n_{p}/n_{i}$ and $\mathbf{v}\approx 
\mathbf{v}_{i}$ (when the mass of electrons and positrons is neglected, and it is also the composite flow velocity of the plasma system). Using
Ampere's law and definition of current density $\mathbf{J}$ ($c\mathbf{
\nabla} \times\mathbf{b}/4\pi =\mathbf{J}$ --the displacement current term is
neglected due to non-relativistic flows of plasma species), the expression
for composite flow velocity $\mathbf{v}\approx \mathbf{v}_{i}$ in normalized
form can be written as
\begin{equation}
\mathbf{v=\nabla \times b}-N_{p}\mathbf{v}_{p}+N_{e}\mathbf{v}_{e}.
\label{an4}
\end{equation}
By substituting the values of $\mathbf{v}_{e}$ and $\mathbf{v}_{p}$ from Eq. (\ref{an4}) in Eqs. (\ref{n1}-\ref{n2}), we obtain the following relation
\begin{equation}
\frac{\partial \mathbf{A}}{\partial t}=\frac{1}{\chi }\left( \mathbf{
v-\nabla \times b}\right) \times \mathbf{b}-\mathbf{\nabla }\left( \chi
\varphi -p_{e}-p_{p}\right),   \label{n5}
\end{equation}
where $\chi =(n_{e}-n_{p})/n_{i}$. At this point, it is of the utmost importance to point out that in order to close the model equations, it is necessary to account for both the continuity equation ($(\partial n_{\alpha}/\partial t) + \mathbf{\nabla}\cdot (n_{\alpha}\mathbf{V}_{\alpha}) =0$) and the equation of state ($p_{\alpha}=n_{\alpha}T_{\alpha}$) for each plasma species. By taking the curl of Eq. (\ref{n5}), we get
\begin{equation}
\frac{\partial \mathbf{b}}{\partial t}=\frac{1}{\chi }\mathbf{\nabla }\times
\left[ \left( \mathbf{v-\nabla \times b}\right) \times \mathbf{b}\right],\label{i1}
\end{equation}
Similarly, by using Eq. (\ref{n5}) in Eq. (\ref{n3}), we get
\begin{equation}
\frac{\partial \mathbf{v}}{\partial t}=\mathbf{v}\times \left( \mathbf{
\nabla \times v}+\mathbf{b}\right) -\frac{1}{\chi }\left( \mathbf{v-\nabla
\times b}\right) \times \mathbf{b}-\mathbf{\nabla }\left( \frac{1}{2}
v^{2}+\frac{z_{d}n_{d}}{n_{i}}\varphi+p\right) ,  \label{v1}
\end{equation}
where $p=p_{i}+p_{e}+p_{p}$. It is important to highlight that the mathematical structure of our model Eqs. (\ref{i1}-\ref{v1}) is analogous to that of HMHD. Since the HMHD is a multi-fluid model that captures the dynamics of both electrons and ions, the Hall term ($(\mathbf{\nabla}\times\mathbf{B})\times\mathbf{B}$) accounts for the electron inertia. In our scenario, the effects of electrons, positrons and dust particles are not completely disregarded. As previously stated, the plasma being studied is quasineutral, and the quasineutrality condition can be stated as  $n_{i}+n_{p}\approx
n_{e}+z_{d}n_{d}$, where $z_{d}$ is the number of electrons residing on the
surface of the dust particle. Then, the Hall term and parameter $\chi$ explicitly account for the impact of these species on the overall dynamics, enabling us to analyze their roles in DY and RDY processes.

Employing the standard methodology outlined in \citeauthor{Mahajan2005} (2005), the generic magnetic field ($\mathbf{b}$) and velocity ($\mathbf{v}$) are decomposed into the equilibrium seed fields ($\mathbf{b}_{0}$ and $\mathbf{v}_{0}$) and the fluctuations. These fluctuations consist of both macroscopic ($\mathbfcal{B}$ and $\mathbfcal{V}$) and microscopic ($\widetilde{\mathbf{b}}$ and $\widetilde{\mathbf{v}}$) components, denoted as
\begin{eqnarray}
    \mathbf{b} =\mathbfcal{B}+\mathbf{b}_{0}+\widetilde{\mathbf{b}},\notag \\
\mathbf{v} =\mathbfcal{V}+\mathbf{v}_{0}+\widetilde{\mathbf{v}}.\label{pte}
\end{eqnarray}
It is crucial to note that in Eq. (\ref{pte}), $\mathbf{b}_{0}$ and $\mathbf{v}_{0}$ represent the steady-state equilibrium solutions to Eqs. (\ref{i1}-\ref{v1}) when there are no macroscopic fluctuations ($\mathbfcal{B}$ and $\mathbfcal{V}$), and these solutions are isotropic and homogeneous. On the other hand, macroscopic fields $\mathbfcal{B}$ and $\mathbfcal{V}$ are macroscopic fluctuations or perturbations that are spatially or statistically averaged, while $\widetilde{\mathbf{b}}$ and $\widetilde{\mathbf{v}}$ are microscopic corrections to the isotropic and homogeneous solutions that are the consequence of the presence of macroscale fields. Consequently, they are not required to be isotropic. Moreover, each of the microscale fields satisfy the condition $\left\langle{\mathbf{b}_{0}}\right\rangle =\left\langle {\mathbf{v}_{0}}\right\rangle=\left\langle{\widetilde{\mathbf{b}}}\right\rangle=\left\langle {\widetilde{\mathbf{v}}}\right\rangle=0$, with the bracket $\left\langle {\cdots}\right\rangle$ denoting an average that satisfies Taylor's hypothesis. However, the products of these microscale fields do not generally satisfy this condition \cite{Mininni2003,Krause1980}. 

The equilibrium fields ($\mathbf{b}_{0}$ and $\mathbf{v}_{0}$) also represent the background turbulence and serve as the energy reservoir that drives the fluctuations. Since these equilibrium fields ($\mathbf{b}_{0}$ and $\mathbf{v}_{0}$) are of a small scale or microscopic--length scale that is smaller than or order of $l_{i}$, generated by some microscopic process, experience saturation, and build up energy that drives both large and small scale fluctuations. Additionally, we can simplify our analysis by considering that the equilibrium fields exist solely at the microscopic scale. We can establish a hierarchy within the microscale fields where the ambient fields significantly surpass the fluctuations at the corresponding scale, such that $\vert\widetilde{\mathbf{b}}\vert\ll\vert\mathbf{b}_{0}\vert$ and $\vert\widetilde{\mathbf{v}}\vert\ll\vert\mathbf{v}_{0}\vert$. For a comprehensive discussion of the closure scheme and its underlying assumptions, see Refs. \cite{Mininni2002,Mininni2003,Mahajan2005}.

At this juncture, it is also imperative to emphasize that the constant-density plasma system is an analytically tractable model that can adequately characterize the unified DY/RDY action. Therefore, for our analysis, we assume a constant density plasma system and choose the Beltrami-Bernoulli class of equilibrium solutions to Eqs. (\ref{i1}-\ref{v1}) for the ambient microscale fields \cite{Lingam2015}. The equilibrium solution to Eqs. (\ref{i1}-\ref{v1}) can be expressed in the following Beltrami-Bernoulli conditions \cite{Mahajan1998}
\begin{equation}
\mathbf{v}_{0}-\mathbf{\nabla \times b}_{0}=\frac{\chi }{a}\mathbf{b}_{0},\label{b1}
\end{equation}
\begin{equation}
\mathbf{\nabla \times v}_{0}+\mathbf{b}_{0}=d\mathbf{v}_{0},\label{b2}
\end{equation}
\begin{equation}
p_{0}+\frac{z_{d}n_{d}}{n_{i}}\varphi_{0}+\frac{1}{2}
v_{0}^{2}=\text{constant},\label{bernoulli}
\end{equation}
where the constants $a$ and $d$, also called Beltrami parameters, are set by the constants of the motion of the equilibrium system, which are the magnetic helicity ($\int \mathbf{A}_{0}\cdot \mathbf{b}_{0}d^{3}x$) and generalized helicity ($\int (\mathbf{v}_{0}+\mathbf{A}_{0})\cdot (\mathbf{\nabla}\times\mathbf{v}_{0}+\mathbf{b}_{0})d^{3}x$). Moreover, in addition to ensuring the homogeneity of plasma energy, Eq. (\ref{bernoulli}) also acts as a closure for the ambient plasma pressure  \cite{Ohsaki2002}. It is also important to note that the steady-state continuity equation is automatically satisfied by the system of Eqs. (\ref{b1}-\ref{bernoulli}) describing the equilibrium state of the plasma system under the incompressibility condition, which ensures the mass conservation of plasma species. By solving Eqs. (\ref{b1}-\ref{b2}), we can obtain the following double Beltrami (DB) equilibrium state for $\mathbf{b}_{0}$ ($\mathbf{v}_{0}$)
\begin{equation}
\mathbf{\nabla} \times \mathbf{\nabla} \times \mathbf{b}_{0}-\left( d-\frac{\chi }{a}\right) 
\mathbf{\nabla} \times \mathbf{b}_{0}+\left( 1-\frac{\chi d}{a}\right) \mathbf{b}
_{0}=0.\label{db}
\end{equation}
The same equilibrium state equation for $\mathbf{b}_{0}$ and $\mathbf{v}_{0}$ also highlights the strong magnetofluid coupling. Further, being the linear superposition of two single Beltrami fields ($\mathbf{\nabla}\times\mathbf{b}_{\pm 0}=\lambda_{\pm}\mathbf{b}_{\pm 0}$), the DB state is characterized by two inverse length scales ($\lambda_{\pm}$). The expressions for the values of the inverse length scales $\lambda_{\pm}$ obtained from DB Eq. (\ref{db}) are as follows
\begin{equation}
\lambda _{\pm }=\frac{1}{2}\left[ \left( d-\frac{\chi }{a}\right) \pm \sqrt{
\left( d+\frac{\chi }{a}\right) ^{2}-4}\right]. \label{isl}
\end{equation}
Proper adjustment of $a$ and $d$ can result in disparate inverse length scales, where $\lambda_{+}=\lambda$ and $\lambda_{-}=\mu$; the $\lambda$ ($\mu$) represents the inverse scale length at a microscopic (macroscopic) scale. In the following discussion, we will adopt $\lambda$ as the macroscopic inverse length scale for the microscopic ambient field and flow ($\mathbf{b}_{0}$ and $\mathbf{v}_{0}$). Therefore, we can express the relation between $\mathbf{v}_{0}$ and $\mathbf{b}_{0}$ as follows
\begin{equation}
\mathbf{v}_{0}=\left( \lambda +\frac{\chi }{a}\right) \mathbf{b}_{0}.
\end{equation}
Additionally, opting for these particular length scales holds significant physical implications. Since the astrophysical systems are on a macroscopic scale, their underlying physics may contain a significant microscopic aspect. Therefore, it is reasonable to assume that the ambient fields are microscopic. This assumption is based on the understanding that the source of larger-scale phenomena is microscopic in nature. After employing the aforementioned assumptions and performing extensive algebraic calculations \cite{Mahajan2005,Lingam2015}, the following equations describing the evolution of the micro- and macroscopic fluctuations are obtained:
\begin{equation}
\frac{\partial \widetilde{\mathbf{v}}}{\partial t}=\left( \frac{1}{\chi }
\mathbfcal{B}-\left( \lambda +\frac{\chi }{a}\right) \mathbfcal{V}\right) \cdot 
\mathbf{\nabla b}_{0}, \label{vm1}
\end{equation}
\begin{equation}
\frac{\partial \widetilde{\mathbf{b}}}{\partial t}=\left( \frac{1}{a}\mathbfcal{
B}-\frac{1}{\chi }\mathbfcal{V}\right) \cdot \mathbf{\nabla b}_{0},\label{bm1}
\end{equation}
\begin{equation}
\frac{\partial ^{2}\mathbfcal{V}}{\partial t^{2}}=\mathbf{\nabla }\times \left(
g\mathbfcal{V}-h\mathbfcal{B}\right),\label{M1} 
\end{equation}
\begin{equation}
\frac{\partial ^{2}\mathbfcal{B}}{\partial t^{2}}=-\mathbf{\nabla }\times
\left( r\mathbfcal{B}+s\mathbfcal{V}\right),\label{M2}
\end{equation}
where
\begin{eqnarray*}
g &=&\frac{\lambda b_{0}^{2}}{6}\left[ \left( \lambda +\frac{\chi }{a}
\right) ^{2}-\frac{1}{\chi ^{2}}\right] , \\
h &=&\frac{\lambda b_{0}^{2}}{6}\left( \frac{1}{a}+\frac{\lambda }{\chi }-
\frac{1}{a\chi }\right) , \\
r &=&\frac{\lambda b_{0}^{2}}{3}\left( \frac{1}{\chi ^{2}}-\frac{\lambda }{
a\chi }-\frac{1}{a^{2}}\right) , \\
s &=&-\frac{\lambda b_{0}^{2}}{3}\left[ \frac{\lambda }{\chi }\left( 1-\frac{
1}{\chi }\right) -\frac{1}{a\chi }\left( 1-\chi \right) \right],
\end{eqnarray*}
in which $b_{0}^{2}$  is ambient microscale magnetic energy. The Eqs. (\ref{vm1}-\ref{M2}) clearly demonstrate that the ambient microscopic dynamics control the macroscopic dynamics of the plasma system, as the microscopic helicities and densities of plasma species determine $\lambda$. By performing the Fourier analysis on Eqs. (\ref{M1}-\ref{M2})  yields the following dispersion relation at which $\mathbfcal{V}$ and $\mathbfcal{B}$ grow:
\begin{equation}
\omega ^{8}-\omega ^{4}k^{2}\left( r^{2}+g^{2}+2hs\right) +k^{4}\left(
gr+hs\right) ^{2}=0.\label{ds}
\end{equation}
From Eq. (\ref{ds}) one can obtain $\omega_{\pm}^{4}=k^{2}\left(g^{2}+r^{2}+2hs\pm \left(g- r\right)\sqrt{(g+r)^{2}+4hs}
\right)/2$. By using the dispersion relation (\ref{ds}), the macroscale fields $\mathbfcal{V}$ and $\mathbfcal{B}$ can be related as
\begin{equation}
\mathbfcal{V}=\frac{h\left(\omega^{4}-(gr+hs)k^{2}\right) }{\left( r\omega^{4}-(gr+hs)gk^{2}\right) }\mathbfcal{B}.\label{MA}
\end{equation} 
At this point, it is also very important to mention that, for our further investigation, we will only use  $\omega^{4}=\omega_{-}^{4}$, while the value of $\omega_{+}^{4}$ is not feasible for our numerical analysis. Since $\mathbfcal{V}$ and $\mathbfcal{B}$ are normalized with $v_{A}$ and $B_{0}$, then the macroscale Alfv\'{e}n Mach number $\mathcal{M}_{A}$ is equal to 
\begin{equation*}
  \mathcal{M}_{A}=\frac{h\omega_{-}^{4}-(gr+hs)hk^{2})}{r\omega_{-}^{4}
-(gr+hs)gk^{2}}.  
\end{equation*}
Similarly, by conducting Fourier analysis on equations (\ref{vm1}-\ref{bm1}) and utilizing the value of $\mathbfcal{V}$ provided by equation (\ref{MA}), we can establish the subsequent relationship between $\widetilde{\mathbf{v}}$ and $\widetilde{\mathbf{b}}$
\begin{equation}
\widetilde{\mathbf{v}}=\widetilde{\mathcal{M}}_{A}\widetilde{\mathbf{b}},\label{ma}
\end{equation}
where microscale Alfv\'{e}n Mach number  $\widetilde{\mathcal{M}}_{A}$ is
\begin{equation*}
  \widetilde{\mathcal{M}}_{A}=\frac{ar(\mathcal{M}_{A} -g^{2})-h\chi(a\lambda+\chi)(\mathcal{M}_{A} -gr)}{r\chi(\mathcal{M}_{A} -g^{2})-ah(\mathcal{M}_{A} -gr)}.  
\end{equation*}
Importantly, Eqs. (\ref{MA}-\ref{ma}) illustrate a unified DY/RDY mechanism--the amplification or generation of macro- and microscale magnetic fields and flows from ambient microscopic turbulent magnetic or kinetic energy. Moreover, for the specified value of $\omega^{4}$, the equilibrium properties of the plasma system are the sole determinant of the evolution of the macro- and microscale fields and flows, regardless of the wave number. Another crucial aspect to note is that, contrary to prior research in the framework of double Beltrami equilibrium states, the evolution of $\mathbfcal{B}$ in Eq. (\ref{M2}) is also dependent on $\mathbfcal{V}$ \cite{Mahajan2005,Lingam2015,Kotorashvili2020}.

\section{Numerical analysis} \label{s3}
In order to analyze the characteristics of the unified DY/RDY mechanism, we focus on the plasma parameters commonly found in AGN plasma. Despite ongoing research, the specific composition of plasma in AGN environments remains uncertain. However, there is a prevalent belief that AGN environments can contain EPID plasma \cite{Rees1984,Lightman1987,Miller1988,Blandford2019,Czerny2023}. For the present study, the plasma density and ambient magnetic field are set to $n_{i} = 10^{10}$ cm$^{-3}$ and $B_{0}=10^{2}$ G \cite{Krishan1997,Asano2007}. It is also explicit from equations (\ref{MA}-\ref{ma}) that the invariant helicities and densities of plasma species play a crucial role in determining the final results, as determined by the Beltrami parameters ($a$, $d$) and $\chi$. Furthermore, disparate scales are present in the astrophysically relevant regime due to the fact that the size of the structure is considerably larger than the ion skin depth. Thus, these constants are chosen to guarantee a large separation between the characteristic scales ($\lambda_{\pm}$). Within this context, our analysis will concentrate on two extremely different situations.

\begin{figure}[h!]
     \centering
     \begin{subfigure}[b]{0.49\textwidth}
         \centering
         \includegraphics[width=\textwidth]{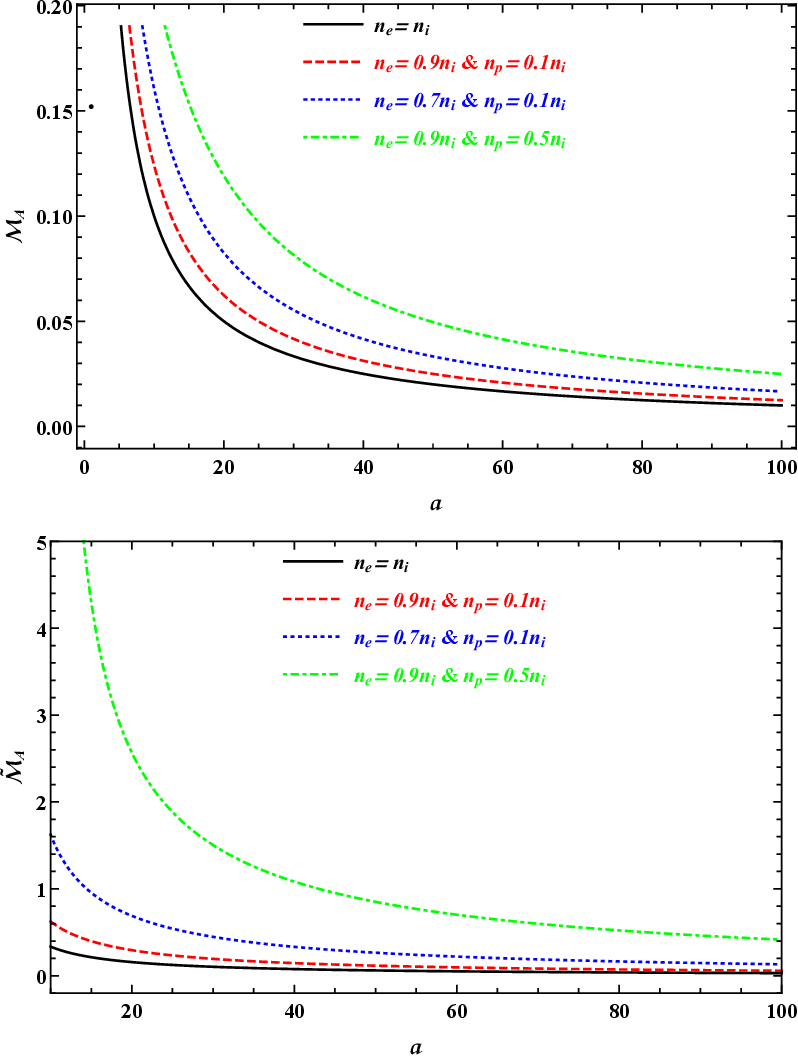}
         \caption{Plot of the Alfv\'{e}n Mach numbers $\mathcal{M}_{A}$ (for macroscale magnetic field $\mathbfcal{B}$ and velocity $\mathbfcal{V}$--top) and $\widetilde{\mathcal{M}}_{A}$ (for microscale magnetic field $\widetilde{\mathbf{b}}$ and velocity $\widetilde{\mathbf{v}}$--bottom) versus $a$ for $a\sim d >1$ and different values of plasma species densities.}
         \label{fig:1a}
     \end{subfigure}
     \hfill
     \begin{subfigure}[b]{0.49\textwidth}
         \centering
         \includegraphics[width=\textwidth]{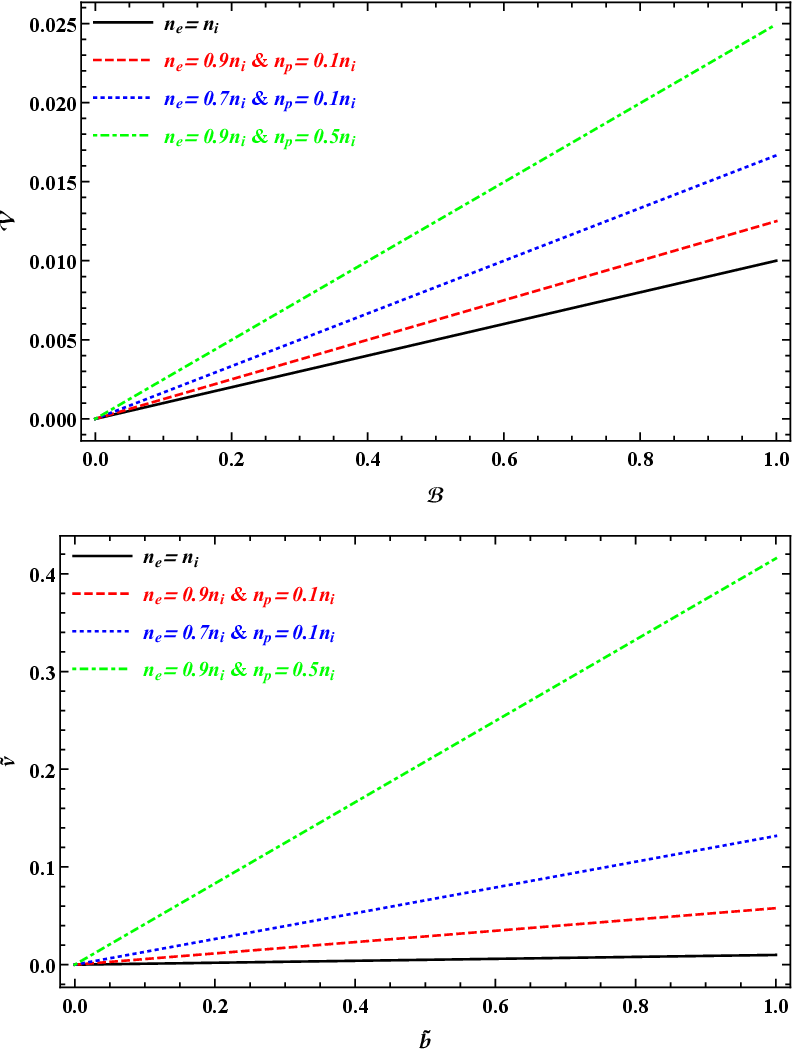}
         \caption{Plot of the macroscale velocity $\mathbfcal{V}$ versus macroscale magnetic field $\mathbfcal{B}$ (top) and microscale velocity $\widetilde{\mathbf{v}}$ versus microscale magnetic field $\widetilde{\mathbf{b}}$ (bottom) for different values of plasma species densities and Beltrami parameters $a\sim d =100$.}
         \label{fig:1b}
     \end{subfigure}
        \caption{Manifestation of the straight DY mechanisms occurring at both macro- and microscales, driven by primarily kinetic microscale ambient fields for different values of plasma species densities.}
        \label{fig:1}
\end{figure}
In the first scenario, we examine a situation where $a\sim d \gg1$, with different values of plasma species densities. In this case, the inverse microscale is $\lambda=\lambda_{+}\gg1$, which demonstrates that $\mathbf{v}_{0}\gg \mathbf{b}_{0}$. This indicates that the microscale fields in the ambient environment are predominantly kinetic. These types of conditions can be found in the plasmas of AGN environments, where the turbulent velocity field may become dominant at certain stages while also having the presence of magnetic field. For instance, highly turbulent flows may exist in the accretion disks of AGN to facilitate the accretion of gas. Also, the broadening of emission line profiles in AGN serves as a sign of the existence of velocity turbulence \cite{Foschini2002}. So as a result of these microscale super-Alfv\'{e}nic flows in the ambient environment, the generated macro- and microscale fields have the exact opposite ordering, i.e., $\mathbfcal{V}=\mathcal{M}_{A}\mathbfcal{B}\ll \mathbfcal{B}$ and $\widetilde{\mathbf{v}}=\widetilde{\mathcal{M}}_{A} \widetilde{\mathbf{b}}\ll\widetilde{\mathbf{b}}$, with respect to these Beltrami parameters. It's worth noting that recent studies have revealed that a notable portion of AGN settings possess remarkably powerful magnetic fields. One possible pathway for evolution involves the amplification of a seed magnetic field through a DY mechanism. Consequently, it is important to demonstrate that the dynamic evolution of the magnetic field via macroscale DY mechanisms may result from the effect of magnetofluid couplings in AGN environments.

Fig. \ref{fig:1} illustrates the impact of an ambient microscale turbulent velocity field for different values of plasma species densities on the evolution of generated macro- and microscale magnetic and velocity fields. In Fig. \ref{fig:1a}, the variations in Alfv\'{e}n Mach numbers $\mathcal{M}_{A}$ (top) and $\widetilde{\mathcal{M}}_{A}$ (bottom) for the generated macroscale and microscale vector fields are shown in relation to the Beltrami parameter $a>1$ (where $a\sim d$); whereas in Fig. \ref{fig:1b}, the plots for the generated macroscale velocity $\mathbfcal{V}$ versus macroscale magnetic field $\mathbfcal{B}$ (top) and the microscale velocity $\widetilde{\mathbf{v}}$ versus microscale magnetic field $\widetilde{\mathbf{b}}$ (bottom) for $a\sim d=100$ are displayed. As the value of the Beltrami parameter increases for the fixed values of plasma species densities, the values of both $\mathcal{M}_{A}$ and $\widetilde{\mathcal{M}}_{A}$ decrease, as shown in Fig. \ref{fig:1a}; for higher values of the Beltrami parameter $a$, $\mathcal{M}_{A}$ and $\widetilde{\mathcal{M}}_{A}$ are both much smaller than 1. Similarly, for the given values of plasma parameters, the generated macroscale and microscale velocities in relation to the corresponding magnetic fields at the macroscale and microscale are both sub-Alfv\'{e}nic, as illustrated in Fig. \ref{fig:1b}. Furthermore, in contrast to prior investigations (where $n_{e}\approx{n}_{i}$) \cite{Mahajan2005,Lingam2015,Kotorashvili2020}, the major finding of the current study is that variations in $\mathcal{M}_{A}$ and $\widetilde{\mathcal{M}}_{A}$, in addition to the velocities generated at the macroscopic and microscale levels, are extremely sensitive to the relative densities of plasma species. From this analysis, it can be inferred that the magnetofluid coupling ensures a straight DY mechanism at both macro- and microscales ($\mathbfcal{B}\gg \mathbfcal{V}$ and $\widetilde{\mathbf{b}}\gg \widetilde{\mathbf{v}}$ ), predominantly originating from super-Alfv\'{e}nic ambient microscale flows ($\mathbf{v}_{0}\gg \mathbf{b}_{0}$). Additionally, the generated flows exhibit sensitivity to the composition of the ambient plasma.

It is noteworthy to emphasize that scientists have identified the potential for generating magnetic fields within the central regions of black hole accretion disks in AGNs \cite{Chakrabarti1994}. In this regard, \citeauthor{Colgate1997} (1997) have also put forth arguments regarding the necessity and feasibility of a strong DY mechanism. Significantly, the variability in polarization also serves as an indication of the presence of strong magnetic fields in AGN environments \cite{Bao1997}. Moreover, Pariev \textit{et al}. suggested that the presence of a magnetic field DY within the inner regions of the accretion disk encompassing the supermassive black holes in AGNs might potentially serve as the underlying process responsible for the generation of magnetic fields in galaxies and extra-galactic space \cite{Pariev2007a,Pariev2007b}. Thus, given the aforementioned discourse, the straight DY mechanism in EPID plasmas will be useful for improving comprehension of the existence and generation of large-scale magnetic fields in AGNs, as well as feedback mechanisms related to these fields.

\begin{figure}[h!]
     \centering
     \begin{subfigure}[b]{0.49\textwidth}
         \centering
         \includegraphics[width=\textwidth]{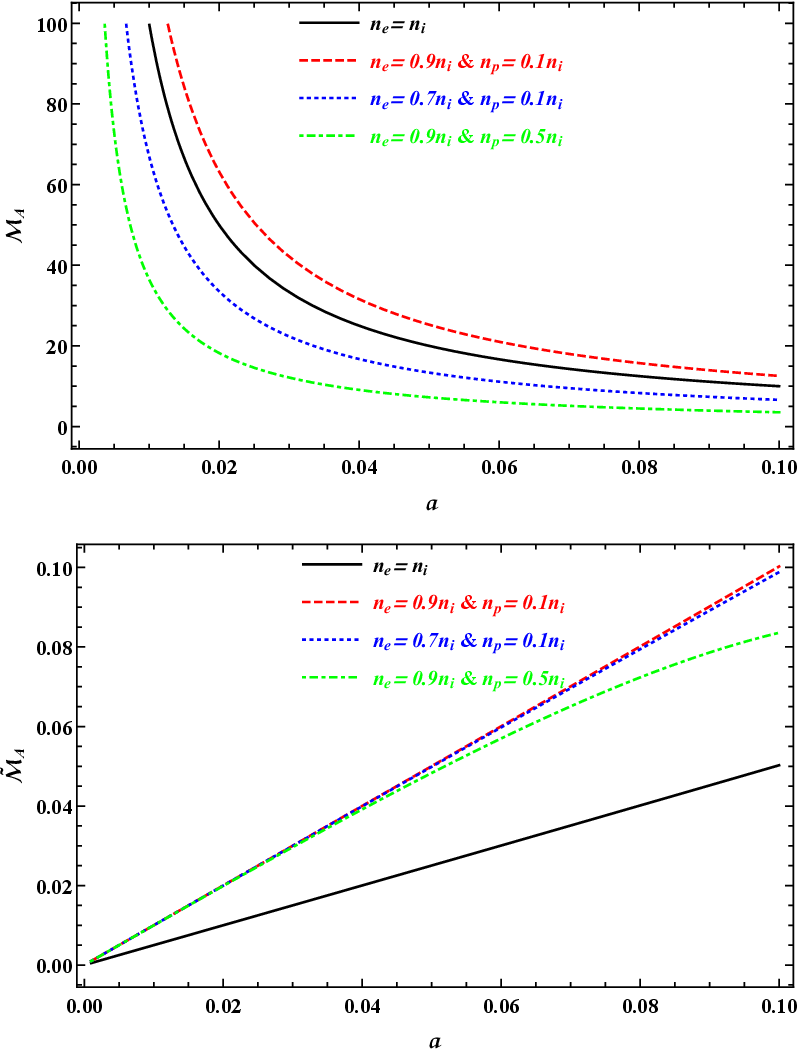}
         \caption{Plot of the Alfv\'{e}n Mach numbers $\mathcal{M}_{A}$ (for macroscale magnetic field $\mathbfcal{B}$ and velocity $\mathbfcal{V}$--top) and $\widetilde{\mathcal{M}}_{A}$ (for microscale magnetic field $\widetilde{\mathbf{b}}$ and velocity $\widetilde{\mathbf{v}}$--bottom) versus $a$ for $a\sim d <1$ and various plasma species densities.}
         \label{fig:2a}
     \end{subfigure}
     \hfill
     \begin{subfigure}[b]{0.49\textwidth}
         \centering
         \includegraphics[width=\textwidth]{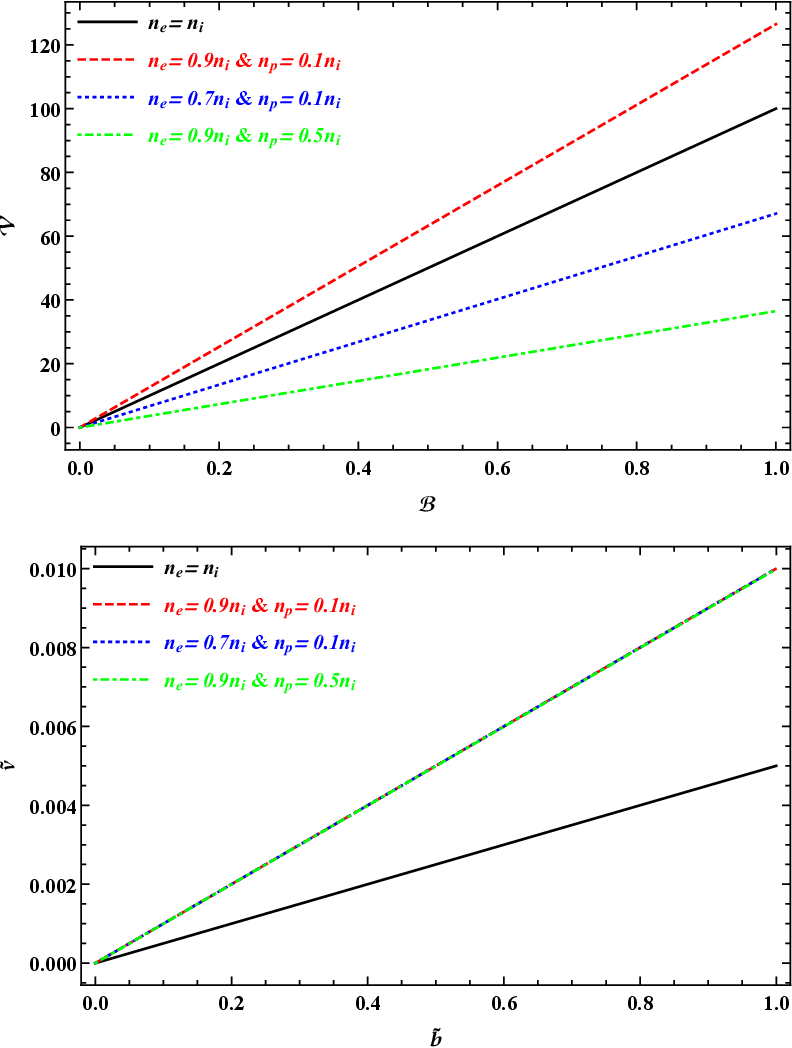}
         \caption{Plot of the macroscale velocity $\mathbfcal{V}$ versus macroscale magnetic field $\mathbfcal{B}$ (top) and microscale velocity $\widetilde{\mathbf{v}}$ versus microscale magnetic field $\widetilde{\mathbf{b}}$ (bottom) for different values of plasma species densities and Beltrami parameters $a\sim d =0.01$.}
         \label{fig:2b}
     \end{subfigure}
        \caption{Manifestation of the unified RDY/DY mechanisms, where the RDY mechanism operates at the macroscale and the DY mechanism operates at the microscale, driven by primarily magnetic microscale ambient fields for different values of plasma species densities.}
        \label{fig:2}
\end{figure}
The second scenario involves analyzing a situation where $a\sim d \ll1$, with different values of plasma species densities. In this situation, the inverse microscale is $\lambda=\lambda_{-}\gg1$, implying that $\mathbf{v}_{0}\ll \mathbf{b}_{0}$. This condition suggests that magnetic fields predominate among the microscale fields present in the ambient environment. These kinds of conditions are present in the plasmas of AGN settings, where the turbulent magnetic field can become dominant at certain stages while exhibiting a strongly sub-Alfv\'{e}nic turbulent flow. In relation to the selected Beltrami parameters, the generated macro and microscale fields exhibit the following ordering: $\mathbfcal{V}=\mathcal{M}_{A}\mathbfcal{B}\gg \mathbfcal{B}$ and $\widetilde{\mathbf{v}}=\widetilde{\mathcal{M}}_{A} \widetilde{\mathbf{b}}\ll\widetilde{\mathbf{b}}$. The obtained conditions suggest the presence of a RDY mechanism at the macroscale and a straight DY mechanism at the microscale, collectively referred to as a unified RDY/DY mechanism. It's interesting to point out that recent studies have uncovered that a significant number of AGN environments exhibit extremely strong outflows and jets. One potential pathway for evolution involves the amplification of a turbulent magnetic field through a unified RDY/DY mechanism. Therefore, it is crucial to showcase the dynamic evolution of the flow and field through a unified RDY/DY mechanism, which can be attributed to the impact of magnetofluid couplings in AGN environments.

The effect of an ambient microscale turbulent magnetic field on the evolution of generated macro- and microscale magnetic and velocity fields is shown in Fig. \ref{fig:2a} for various values of plasma species densities. The variations in the Alfv\'{e}n $\mathcal{M}_{A}$ (top) and and $\widetilde{\mathcal{M}}_{A}$ (bottom) of the generated macroscale and microscale vector fields with respect to the Beltrami parameter $a<1$ (where $a\sim d$) are illustrated in Figure \ref{fig:2a}. The plots of the generated macroscale velocity $\mathbfcal{V}$ versus macroscale magnetic field $\mathbfcal{B}$ (top) and the microscale velocity $\widetilde{\mathbf{v}}$ versus microscale magnetic field $\widetilde{\mathbf{b}}$ (bottom) are presented in Figure \ref{fig:2b}, where the values of Beltrami parameters are $a\sim d=0.01$. As can be seen in Fig. \ref{fig:2a}, the values of $\mathcal{M}_{A}$ are extremely high and significantly larger than unity, while the values of $\widetilde{\mathcal{M}}_{A}$ are significantly smaller than unity. Furthermore, it is observed that the values of both $\mathcal{M}_{A}$ and $\widetilde{\mathcal{M}}_{A}$ decrease as the value of the Beltrami parameter $a$ increases for the fixed values of plasma species densities. From Fig. \ref{fig:2b}, it is also very clear that the generated macro- and microscale velocities in proportion to the associated magnetic fields are super-Alfv\'{e}nic and sub-Alfv\'{e}nic, respectively, for the given values of plasma parameters.
Furthermore, it is important to note that even in this specific case, in comparison to previous investigations (where $n_{e}\approx{n}_{i}$) \cite{Mahajan2005,Lingam2015,Kotorashvili2020}, the variations in $\mathcal{M}_{A}$ and $\widetilde{\mathcal{M}}_{A}$, as well as velocities generated at the macro- and microscales, are highly sensitive to the relative densities of plasma species. Based on this analysis, it can be deduced that the magnetofluid coupling guarantees a unified RDY/DY mechanism ($\mathbfcal{V}\gg \mathbfcal{B}$ and $\widetilde{\mathbf{v}}\ll \widetilde{\mathbf{b}}$ ), primarily resulting from the dominant magnetic fields at the ambient microscale with sub-Alfv\'{e}nic ambient microscale flows ($\mathbf{b}_{0}\gg \mathbf{v}_{0}$). Additionally, the strength of the generated flows and
fields are also dependent on the composition of the plasma that exists in the ambient environment.

It is important to highlight that around 50\% of AGNs are found to have
blueshifted absorption lines for ultraviolet and X-rays, this indicates the widespread prevalence of outflows in AGNs \cite{Crenshaw2003}. Also, the absorption characteristics of the AGN X-ray spectra are particularly intriguing due to their ability to cover a broader spectrum of ionization states. Due to the higher ionization states of their plasma, they make it impossible for winds to be generated by line radiation pressure. These facts support the notion that MHD-driven winds are more favorable \cite{Blandford1982,Contopoulos1994}. The self-similar structure of these winds also allows them to cover a wide range of radii. Furthermore, these winds are not dependent on radiation pressure for their initiation. Interestingly, the investigation of the polarization of dust emission in Cygnus A provides further evidence in favor of the wind structure that is primarily influenced by magnetic fields \cite{Lopez2018}. In a study by \citeauthor{Vlahakis2004} (2004), it was also shown that the acceleration of plasma flows to relativistic speeds observed in sources like the radio galaxy NGC 6251 and the quasar 3C 345 can be explained by magnetic driving. Since plasma acceleration, outflows, and jets are frequently observed in AGN environments, and their origin is still a topic of continued debate. Also the previous literature suggest that fast plasma flows are often driven by magnetic energy \cite{Crenshaw2003,Blandford1982,Contopoulos1994,Lopez2018,Vlahakis2004}. As the findings of the present investigation highlight the generation of fast plasma flows through a unified RDY/DY mechanism driven by microscale turbulent magnetic energy. Therefore, the present study can aid in gaining a deeper understanding of this plasma phenomenon in astrophysical settings.
\section{Summary} \label{s4}
In the present study, the possibility of the unified DY/RDY mechanism, which is a manifestation of magnetofluid coupling, has been investigated in a four-component astrophysical dusty plasma. The unified DY/RDY mechanism is a phenomenon that involves the simultaneous generation of a macroscale magnetic field and flow from ambient microscale turbulent magnetic or kinetic energy. We have considered a magnetized, quasineutral and incompressible four-component plasma in our theoretical model, which consists of mobile massless electrons and positrons, inertial positively charged singly ionized ions and negatively charged static dust particles. Using the model equations for dynamic plasma species and within the framework of the double Beltrami equilibrium state for ambient microscale magnetic and velocity fields, we have derived the evolution equations for the generated macro- and microscale magnetic and velocity fields. Unlike previous studies that focused on two-fluid (electron-ion) plasmas with double Beltrami equilibrium states \cite{Mahajan2005,Lingam2015,Kotorashvili2020}, it is important to highlight that both macroscale magnetic fields and flow control the evolution of macroscale magnetic fields. From the macroscopic evolution equations for fields and flows, an equation for a unified DY/RDY mechanism and dispersion relation have been derived. In a similar vein, the mathematical relations between generated microscale fields and flows have also been derived. The mathematical expression for the unified DY/RDY mechanism is entirely dependent on plasma species densities and the inverse microscale. The value of inverse-microscale has been determined by invoking the Beltrami-Bernoulli equilibrium states of the plasma system, and it depends on plasma species densities and Beltrami parameters—a measure of invariant helicities. The numerical analysis, which takes into account the plasma parameters characteristic of the AGN environment, reveals that in cases where the predominant turbulent energy at the microscale is kinetic, a straight DY mechanism manifests at both the macroscopic and microscales. Moreover, the Alfv\'{e}n Mach numbers for the generated flows at both the macroscopic and microscales are significantly smaller than unity ($\mathcal{M}_{A}\ll1$ and $\widetilde{\mathcal{M}}_{A}\ll1$). This straight DY mechanism, driven by the ambient short-scale turbulent flows, has the ability to generate a strong macroscale magnetic field. However, the numerical analysis indicates that when the primary turbulent energy at the microscale is magnetic, a RDY mechanism emerges at the macroscale, while a straight DY occurs at the microscale. The aforementioned scenario is commonly known as a unified RDY/DY mechanism. For the generated macroscale flows, $\mathcal{M}_{A}\gg1$, while for microscale flows, $\widetilde{\mathcal{M}}_{A}\ll1$. Therefore, when ambient short-scale magnetic turbulence drives the plasma system, this unified RDY/DY mechanism can generate macroscale fast flows as well as a weak magnetic field. Furthermore, it is important to note that the Alfv\'{e}n Mach numbers for both macro- and microscale flows are also considerably influenced by the relative density of plasma species in both scenarios--straight DY and unified RDY/DY processes. Compared to previous investigations pertaining to two fluid electron-ion plasma in the framework of HMHD \cite{Mahajan2005,Lingam2015,Kotorashvili2020}, the present analysis also reveals that even a minor presence of dust and positron species significantly impacts the generated fields and flows. Due to the fact that the origins of large-scale magnetic fields and fast plasma flows (outflows and jets) in AGN environments are still being discussed, the findings of the current study may be helpful in gaining a better understanding of these phenomena. Additionally, the generated macroscale magnetic fields and fast flows may have an impact on the AGN feedback.
\section*{Data availability statement}
No new data were created or analyzed in this study.
\bibliographystyle{aipnum4-1.bst}
\bibliography{ps}

\begin{thebibliography}{66}%
\makeatletter
\providecommand \@ifxundefined [1]{%
 \@ifx{#1\undefined}
}%
\providecommand \@ifnum [1]{%
 \ifnum #1\expandafter \@firstoftwo
 \else \expandafter \@secondoftwo
 \fi
}%
\providecommand \@ifx [1]{%
 \ifx #1\expandafter \@firstoftwo
 \else \expandafter \@secondoftwo
 \fi
}%
\providecommand \natexlab [1]{#1}%
\providecommand \enquote  [1]{``#1''}%
\providecommand \bibnamefont  [1]{#1}%
\providecommand \bibfnamefont [1]{#1}%
\providecommand \citenamefont [1]{#1}%
\providecommand \href@noop [0]{\@secondoftwo}%
\providecommand \href [0]{\begingroup \@sanitize@url \@href}%
\providecommand \@href[1]{\@@startlink{#1}\@@href}%
\providecommand \@@href[1]{\endgroup#1\@@endlink}%
\providecommand \@sanitize@url [0]{\catcode `\\12\catcode `\$12\catcode `\&12\catcode `\#12\catcode `\^12\catcode `\_12\catcode `\%12\relax}%
\providecommand \@@startlink[1]{}%
\providecommand \@@endlink[0]{}%
\providecommand \url  [0]{\begingroup\@sanitize@url \@url }%
\providecommand \@url [1]{\endgroup\@href {#1}{\urlprefix }}%
\providecommand \urlprefix  [0]{URL }%
\providecommand \Eprint [0]{\href }%
\providecommand \doibase [0]{http://dx.doi.org/}%
\providecommand \selectlanguage [0]{\@gobble}%
\providecommand \bibinfo  [0]{\@secondoftwo}%
\providecommand \bibfield  [0]{\@secondoftwo}%
\providecommand \translation [1]{[#1]}%
\providecommand \BibitemOpen [0]{}%
\providecommand \bibitemStop [0]{}%
\providecommand \bibitemNoStop [0]{.\EOS\space}%
\providecommand \EOS [0]{\spacefactor3000\relax}%
\providecommand \BibitemShut  [1]{\csname bibitem#1\endcsname}%
\let\auto@bib@innerbib\@empty
\bibitem [{\citenamefont {Moffatt}(1978)}]{Moffat1978}%
  \BibitemOpen
  \bibfield  {author} {\bibinfo {author} {\bibfnamefont {H.~K.}\ \bibnamefont {Moffatt}},\ }\href {https://www.damtp.cam.ac.uk/user/hkm2/PDFs/Moffatt1978.pdf} {\emph {\bibinfo {title} {{Magnetic field generation in electrically conducting fluids}}}},\ \bibinfo {edition} {1st}\ ed.\ (\bibinfo  {publisher} {Cambridge University Press},\ \bibinfo {address} {Cambridge},\ \bibinfo {year} {1978})\BibitemShut {NoStop}%
\bibitem [{\citenamefont {Bodin}(1990)}]{Bodin1990}%
  \BibitemOpen
  \bibfield  {author} {\bibinfo {author} {\bibfnamefont {H.}~\bibnamefont {Bodin}},\ }\href {\doibase 10.1088/0029-5515/30/9/005} {\bibfield  {journal} {\bibinfo  {journal} {Nucl. Fusion}\ }\textbf {\bibinfo {volume} {30}},\ \bibinfo {pages} {1717} (\bibinfo {year} {1990})}\BibitemShut {NoStop}%
\bibitem [{\citenamefont {Chahine}, \citenamefont {Bos},\ and\ \citenamefont {Plihon}(2022)}]{Chahine2022}%
  \BibitemOpen
  \bibfield  {author} {\bibinfo {author} {\bibfnamefont {R.}~\bibnamefont {Chahine}}, \bibinfo {author} {\bibfnamefont {W.~J.~T.}\ \bibnamefont {Bos}}, \ and\ \bibinfo {author} {\bibfnamefont {N.}~\bibnamefont {Plihon}},\ }\href {\doibase 10.1063/5.0078860} {\bibfield  {journal} {\bibinfo  {journal} {Phys. Plasmas}\ }\textbf {\bibinfo {volume} {29}},\ \bibinfo {pages} {032306} (\bibinfo {year} {2022})}\BibitemShut {NoStop}%
\bibitem [{\citenamefont {Zweibel}\ and\ \citenamefont {Heiles}(1997)}]{Zweibel1997}%
  \BibitemOpen
  \bibfield  {author} {\bibinfo {author} {\bibfnamefont {E.~G.}\ \bibnamefont {Zweibel}}\ and\ \bibinfo {author} {\bibfnamefont {C.}~\bibnamefont {Heiles}},\ }\href {\doibase 10.1038/385131a0} {\bibfield  {journal} {\bibinfo  {journal} {Nature}\ }\textbf {\bibinfo {volume} {385}},\ \bibinfo {pages} {131} (\bibinfo {year} {1997})}\BibitemShut {NoStop}%
\bibitem [{\citenamefont {Brandenburg}\ and\ \citenamefont {Subramanian}(2005)}]{Brandenburg2005}%
  \BibitemOpen
  \bibfield  {author} {\bibinfo {author} {\bibfnamefont {A.}~\bibnamefont {Brandenburg}}\ and\ \bibinfo {author} {\bibfnamefont {K.}~\bibnamefont {Subramanian}},\ }\href {\doibase 10.1016/j.physrep.2005.06.005} {\bibfield  {journal} {\bibinfo  {journal} {Phys. Rep.}\ }\textbf {\bibinfo {volume} {417}},\ \bibinfo {pages} {1} (\bibinfo {year} {2005})}\BibitemShut {NoStop}%
\bibitem [{\citenamefont {Kulsrud}\ and\ \citenamefont {Zweibel}(2008)}]{Kulsrud2008}%
  \BibitemOpen
  \bibfield  {author} {\bibinfo {author} {\bibfnamefont {R.~M.}\ \bibnamefont {Kulsrud}}\ and\ \bibinfo {author} {\bibfnamefont {E.~G.}\ \bibnamefont {Zweibel}},\ }\href {\doibase 10.1088/0034-4885/71/4/046901} {\bibfield  {journal} {\bibinfo  {journal} {Reports Prog. Phys.}\ }\textbf {\bibinfo {volume} {71}},\ \bibinfo {pages} {046901} (\bibinfo {year} {2008})}\BibitemShut {NoStop}%
\bibitem [{\citenamefont {Durrer}\ and\ \citenamefont {Neronov}(2013)}]{Durrer2013}%
  \BibitemOpen
  \bibfield  {author} {\bibinfo {author} {\bibfnamefont {R.}~\bibnamefont {Durrer}}\ and\ \bibinfo {author} {\bibfnamefont {A.}~\bibnamefont {Neronov}},\ }\href {\doibase 10.1007/s00159-013-0062-7} {\bibfield  {journal} {\bibinfo  {journal} {Astron. Astrophys. Rev.}\ }\textbf {\bibinfo {volume} {21}},\ \bibinfo {pages} {62} (\bibinfo {year} {2013})}\BibitemShut {NoStop}%
\bibitem [{\citenamefont {Roberts}\ and\ \citenamefont {King}(2013)}]{Roberts2013}%
  \BibitemOpen
  \bibfield  {author} {\bibinfo {author} {\bibfnamefont {P.~H.}\ \bibnamefont {Roberts}}\ and\ \bibinfo {author} {\bibfnamefont {E.~M.}\ \bibnamefont {King}},\ }\href {\doibase 10.1088/0034-4885/76/9/096801} {\bibfield  {journal} {\bibinfo  {journal} {Reports Prog. Phys.}\ }\textbf {\bibinfo {volume} {76}},\ \bibinfo {pages} {096801} (\bibinfo {year} {2013})}\BibitemShut {NoStop}%
\bibitem [{\citenamefont {Helander}, \citenamefont {Strumik},\ and\ \citenamefont {Schekochihin}(2016)}]{Helander2016}%
  \BibitemOpen
  \bibfield  {author} {\bibinfo {author} {\bibfnamefont {P.}~\bibnamefont {Helander}}, \bibinfo {author} {\bibfnamefont {M.}~\bibnamefont {Strumik}}, \ and\ \bibinfo {author} {\bibfnamefont {A.~A.}\ \bibnamefont {Schekochihin}},\ }\href {\doibase 10.1017/S0022377816000982} {\bibfield  {journal} {\bibinfo  {journal} {J. Plasma Phys.}\ }\textbf {\bibinfo {volume} {82}},\ \bibinfo {pages} {905820601} (\bibinfo {year} {2016})}\BibitemShut {NoStop}%
\bibitem [{\citenamefont {Rincon}(2019)}]{Rincon2019}%
  \BibitemOpen
  \bibfield  {author} {\bibinfo {author} {\bibfnamefont {F.}~\bibnamefont {Rincon}},\ }\href {\doibase 10.1017/S0022377819000539} {\bibfield  {journal} {\bibinfo  {journal} {J. Plasma Phys.}\ }\textbf {\bibinfo {volume} {85}},\ \bibinfo {pages} {205850401} (\bibinfo {year} {2019})}\BibitemShut {NoStop}%
\bibitem [{\citenamefont {Mininni}, \citenamefont {G{\'{o}}mez},\ and\ \citenamefont {Mahajan}(2002)}]{Mininni2002}%
  \BibitemOpen
  \bibfield  {author} {\bibinfo {author} {\bibfnamefont {P.~D.}\ \bibnamefont {Mininni}}, \bibinfo {author} {\bibfnamefont {D.~O.}\ \bibnamefont {G{\'{o}}mez}}, \ and\ \bibinfo {author} {\bibfnamefont {S.~M.}\ \bibnamefont {Mahajan}},\ }\href {\doibase 10.1086/339850} {\bibfield  {journal} {\bibinfo  {journal} {Astrophys. J.}\ }\textbf {\bibinfo {volume} {567}},\ \bibinfo {pages} {L81} (\bibinfo {year} {2002})}\BibitemShut {NoStop}%
\bibitem [{\citenamefont {Mininni}, \citenamefont {G{\'{o}}mez},\ and\ \citenamefont {Mahajan}(2003{\natexlab{a}})}]{Mininni2003}%
  \BibitemOpen
  \bibfield  {author} {\bibinfo {author} {\bibfnamefont {P.~D.}\ \bibnamefont {Mininni}}, \bibinfo {author} {\bibfnamefont {D.~O.}\ \bibnamefont {G{\'{o}}mez}}, \ and\ \bibinfo {author} {\bibfnamefont {S.~M.}\ \bibnamefont {Mahajan}},\ }\href {\doibase 10.1086/345777} {\bibfield  {journal} {\bibinfo  {journal} {Astrophys. J.}\ }\textbf {\bibinfo {volume} {584}},\ \bibinfo {pages} {1120} (\bibinfo {year} {2003}{\natexlab{a}})}\BibitemShut {NoStop}%
\bibitem [{\citenamefont {Mininni}, \citenamefont {G{\'{o}}mez},\ and\ \citenamefont {Mahajan}(2003{\natexlab{b}})}]{Mininni2003a}%
  \BibitemOpen
  \bibfield  {author} {\bibinfo {author} {\bibfnamefont {P.~D.}\ \bibnamefont {Mininni}}, \bibinfo {author} {\bibfnamefont {D.~O.}\ \bibnamefont {G{\'{o}}mez}}, \ and\ \bibinfo {author} {\bibfnamefont {S.~M.}\ \bibnamefont {Mahajan}},\ }\href {\doibase 10.1086/368181} {\bibfield  {journal} {\bibinfo  {journal} {Astrophys. J.}\ }\textbf {\bibinfo {volume} {587}},\ \bibinfo {pages} {472} (\bibinfo {year} {2003}{\natexlab{b}})}\BibitemShut {NoStop}%
\bibitem [{\citenamefont {{Bellot Rubio}}\ \emph {et~al.}(2001)\citenamefont {{Bellot Rubio}}, \citenamefont {{Rodriguez Hidalgo}}, \citenamefont {Collados}, \citenamefont {Khomenko},\ and\ \citenamefont {{Ruiz Cobo}}}]{Bellot2001}%
  \BibitemOpen
  \bibfield  {author} {\bibinfo {author} {\bibfnamefont {L.~R.}\ \bibnamefont {{Bellot Rubio}}}, \bibinfo {author} {\bibfnamefont {I.}~\bibnamefont {{Rodriguez Hidalgo}}}, \bibinfo {author} {\bibfnamefont {M.}~\bibnamefont {Collados}}, \bibinfo {author} {\bibfnamefont {E.}~\bibnamefont {Khomenko}}, \ and\ \bibinfo {author} {\bibfnamefont {B.}~\bibnamefont {{Ruiz Cobo}}},\ }\href {\doibase 10.1086/323063} {\bibfield  {journal} {\bibinfo  {journal} {Astrophys. J.}\ }\textbf {\bibinfo {volume} {560}},\ \bibinfo {pages} {1010} (\bibinfo {year} {2001})}\BibitemShut {NoStop}%
\bibitem [{\citenamefont {Blackman}\ and\ \citenamefont {Field}(2004)}]{Blackman2004}%
  \BibitemOpen
  \bibfield  {author} {\bibinfo {author} {\bibfnamefont {E.~G.}\ \bibnamefont {Blackman}}\ and\ \bibinfo {author} {\bibfnamefont {G.~B.}\ \bibnamefont {Field}},\ }\href {\doibase 10.1063/1.1739236} {\bibfield  {journal} {\bibinfo  {journal} {Phys. Plasmas}\ }\textbf {\bibinfo {volume} {11}},\ \bibinfo {pages} {3264} (\bibinfo {year} {2004})}\BibitemShut {NoStop}%
\bibitem [{\citenamefont {Socas-Navarro}\ and\ \citenamefont {{Manso Sainz}}(2005)}]{Socas2005}%
  \BibitemOpen
  \bibfield  {author} {\bibinfo {author} {\bibfnamefont {H.}~\bibnamefont {Socas-Navarro}}\ and\ \bibinfo {author} {\bibfnamefont {R.}~\bibnamefont {{Manso Sainz}}},\ }\href {\doibase 10.1086/428397} {\bibfield  {journal} {\bibinfo  {journal} {Astrophys. J.}\ }\textbf {\bibinfo {volume} {620}},\ \bibinfo {pages} {L71} (\bibinfo {year} {2005})}\BibitemShut {NoStop}%
\bibitem [{\citenamefont {Mahajan}\ \emph {et~al.}(2005)\citenamefont {Mahajan}, \citenamefont {Shatashvili}, \citenamefont {Mikeladze},\ and\ \citenamefont {Sigua}}]{Mahajan2005}%
  \BibitemOpen
  \bibfield  {author} {\bibinfo {author} {\bibfnamefont {S.~M.}\ \bibnamefont {Mahajan}}, \bibinfo {author} {\bibfnamefont {N.~L.}\ \bibnamefont {Shatashvili}}, \bibinfo {author} {\bibfnamefont {S.~V.}\ \bibnamefont {Mikeladze}}, \ and\ \bibinfo {author} {\bibfnamefont {K.~I.}\ \bibnamefont {Sigua}},\ }\href {\doibase 10.1086/432867} {\bibfield  {journal} {\bibinfo  {journal} {Astrophys. J.}\ }\textbf {\bibinfo {volume} {634}},\ \bibinfo {pages} {419} (\bibinfo {year} {2005})}\BibitemShut {NoStop}%
\bibitem [{\citenamefont {Lingam}\ and\ \citenamefont {Mahajan}(2015)}]{Lingam2015}%
  \BibitemOpen
  \bibfield  {author} {\bibinfo {author} {\bibfnamefont {M.}~\bibnamefont {Lingam}}\ and\ \bibinfo {author} {\bibfnamefont {S.~M.}\ \bibnamefont {Mahajan}},\ }\href {\doibase 10.1093/mnrasl/slv017} {\bibfield  {journal} {\bibinfo  {journal} {Mon. Not. R. Astron. Soc. Lett.}\ }\textbf {\bibinfo {volume} {449}},\ \bibinfo {pages} {L36} (\bibinfo {year} {2015})}\BibitemShut {NoStop}%
\bibitem [{\citenamefont {Brandenburg}\ and\ \citenamefont {Rempel}(2019)}]{Brandenburg2019}%
  \BibitemOpen
  \bibfield  {author} {\bibinfo {author} {\bibfnamefont {A.}~\bibnamefont {Brandenburg}}\ and\ \bibinfo {author} {\bibfnamefont {M.}~\bibnamefont {Rempel}},\ }\href {\doibase 10.3847/1538-4357/ab24bd} {\bibfield  {journal} {\bibinfo  {journal} {Astrophys. J.}\ }\textbf {\bibinfo {volume} {879}},\ \bibinfo {pages} {57} (\bibinfo {year} {2019})}\BibitemShut {NoStop}%
\bibitem [{\citenamefont {Kotorashvili}, \citenamefont {Revazashvili},\ and\ \citenamefont {Shatashvili}(2020)}]{Kotorashvili2020}%
  \BibitemOpen
  \bibfield  {author} {\bibinfo {author} {\bibfnamefont {K.}~\bibnamefont {Kotorashvili}}, \bibinfo {author} {\bibfnamefont {N.}~\bibnamefont {Revazashvili}}, \ and\ \bibinfo {author} {\bibfnamefont {N.~L.}\ \bibnamefont {Shatashvili}},\ }\href {\doibase 10.1007/s10509-020-03871-w} {\bibfield  {journal} {\bibinfo  {journal} {Astrophys. Space Sci.}\ }\textbf {\bibinfo {volume} {365}},\ \bibinfo {pages} {175} (\bibinfo {year} {2020})}\BibitemShut {NoStop}%
\bibitem [{\citenamefont {Kotorashvili}\ and\ \citenamefont {Shatashvili}(2022)}]{Kotorashvili2022}%
  \BibitemOpen
  \bibfield  {author} {\bibinfo {author} {\bibfnamefont {K.}~\bibnamefont {Kotorashvili}}\ and\ \bibinfo {author} {\bibfnamefont {N.~L.}\ \bibnamefont {Shatashvili}},\ }\href {\doibase 10.1007/s10509-021-04034-1} {\bibfield  {journal} {\bibinfo  {journal} {Astrophys. Space Sci.}\ }\textbf {\bibinfo {volume} {367}},\ \bibinfo {pages} {2} (\bibinfo {year} {2022})}\BibitemShut {NoStop}%
\bibitem [{\citenamefont {Murphy}\ \emph {et~al.}(2005)\citenamefont {Murphy}, \citenamefont {Share}, \citenamefont {Skibo},\ and\ \citenamefont {Kozlovsky}}]{Murphy2005}%
  \BibitemOpen
  \bibfield  {author} {\bibinfo {author} {\bibfnamefont {R.~J.}\ \bibnamefont {Murphy}}, \bibinfo {author} {\bibfnamefont {G.~H.}\ \bibnamefont {Share}}, \bibinfo {author} {\bibfnamefont {J.~G.}\ \bibnamefont {Skibo}}, \ and\ \bibinfo {author} {\bibfnamefont {B.}~\bibnamefont {Kozlovsky}},\ }\href {\doibase 10.1086/452634} {\bibfield  {journal} {\bibinfo  {journal} {Astrophys. J. Suppl. Ser.}\ }\textbf {\bibinfo {volume} {161}},\ \bibinfo {pages} {495} (\bibinfo {year} {2005})}\BibitemShut {NoStop}%
\bibitem [{\citenamefont {Shukla}\ and\ \citenamefont {Mamun}(2002)}]{Shukla2002}%
  \BibitemOpen
  \bibfield  {author} {\bibinfo {author} {\bibfnamefont {P.~K.}\ \bibnamefont {Shukla}}\ and\ \bibinfo {author} {\bibfnamefont {A.~A.}\ \bibnamefont {Mamun}},\ }\href {\doibase https://doi.org/10.1201/9781420034103} {\emph {\bibinfo {title} {{Introduction to Dusty Plasma Physics}}}},\ \bibinfo {edition} {1st}\ ed.\ (\bibinfo  {publisher} {CRC Press},\ \bibinfo {address} {Boca Raton},\ \bibinfo {year} {2002})\BibitemShut {NoStop}%
\bibitem [{\citenamefont {Zurek}(1985)}]{Zurek1985}%
  \BibitemOpen
  \bibfield  {author} {\bibinfo {author} {\bibfnamefont {W.~H.}\ \bibnamefont {Zurek}},\ }\href {\doibase 10.1086/162921} {\bibfield  {journal} {\bibinfo  {journal} {Astrophys. J.}\ }\textbf {\bibinfo {volume} {289}},\ \bibinfo {pages} {603} (\bibinfo {year} {1985})}\BibitemShut {NoStop}%
\bibitem [{\citenamefont {Rees}(1984)}]{Rees1984}%
  \BibitemOpen
  \bibfield  {author} {\bibinfo {author} {\bibfnamefont {M.~J.}\ \bibnamefont {Rees}},\ }\href {\doibase 10.1146/annurev.aa.22.090184.002351} {\bibfield  {journal} {\bibinfo  {journal} {Annu. Rev. Astron. Astrophys.}\ }\textbf {\bibinfo {volume} {22}},\ \bibinfo {pages} {471} (\bibinfo {year} {1984})}\BibitemShut {NoStop}%
\bibitem [{\citenamefont {Lightman}\ and\ \citenamefont {Zdziarski}(1987)}]{Lightman1987}%
  \BibitemOpen
  \bibfield  {author} {\bibinfo {author} {\bibfnamefont {A.~P.}\ \bibnamefont {Lightman}}\ and\ \bibinfo {author} {\bibfnamefont {A.~A.}\ \bibnamefont {Zdziarski}},\ }\href {\doibase 10.1086/165485} {\bibfield  {journal} {\bibinfo  {journal} {Astrophys. J.}\ }\textbf {\bibinfo {volume} {319}},\ \bibinfo {pages} {643} (\bibinfo {year} {1987})}\BibitemShut {NoStop}%
\bibitem [{\citenamefont {Miller}\ and\ \citenamefont {Wiita}(1988)}]{Miller1988}%
  \BibitemOpen
  \bibinfo {editor} {\bibfnamefont {H.~R.}\ \bibnamefont {Miller}}\ and\ \bibinfo {editor} {\bibfnamefont {P.~J.}\ \bibnamefont {Wiita}},\ eds.,\ \href {\doibase 10.1007/3-540-19492-4} {\emph {\bibinfo {title} {{Active Galactic Nuclei}}}}\ (\bibinfo  {publisher} {Springer Berlin},\ \bibinfo {address} {Heidelberg},\ \bibinfo {year} {1988})\BibitemShut {NoStop}%
\bibitem [{\citenamefont {Blandford}, \citenamefont {Meier},\ and\ \citenamefont {Readhead}(2019)}]{Blandford2019}%
  \BibitemOpen
  \bibfield  {author} {\bibinfo {author} {\bibfnamefont {R.}~\bibnamefont {Blandford}}, \bibinfo {author} {\bibfnamefont {D.}~\bibnamefont {Meier}}, \ and\ \bibinfo {author} {\bibfnamefont {A.}~\bibnamefont {Readhead}},\ }\href {\doibase 10.1146/annurev-astro-081817-051948} {\bibfield  {journal} {\bibinfo  {journal} {Annu. Rev. Astron. Astrophys.}\ }\textbf {\bibinfo {volume} {57}},\ \bibinfo {pages} {467} (\bibinfo {year} {2019})}\BibitemShut {NoStop}%
\bibitem [{\citenamefont {Czerny}\ \emph {et~al.}(2023)\citenamefont {Czerny}, \citenamefont {Zaja{\v{c}}ek}, \citenamefont {Naddaf}, \citenamefont {Sniegowska}, \citenamefont {Panda}, \citenamefont {R{\'{o}}{\.{z}}anska}, \citenamefont {Adhikari}, \citenamefont {Pandey}, \citenamefont {Jaiswal}, \citenamefont {Karas}, \citenamefont {Borkar}, \citenamefont {Mart{\'{i}}nez-Aldama},\ and\ \citenamefont {Prince}}]{Czerny2023}%
  \BibitemOpen
  \bibfield  {author} {\bibinfo {author} {\bibfnamefont {B.}~\bibnamefont {Czerny}}, \bibinfo {author} {\bibfnamefont {M.}~\bibnamefont {Zaja{\v{c}}ek}}, \bibinfo {author} {\bibfnamefont {M.-H.}\ \bibnamefont {Naddaf}}, \bibinfo {author} {\bibfnamefont {M.}~\bibnamefont {Sniegowska}}, \bibinfo {author} {\bibfnamefont {S.}~\bibnamefont {Panda}}, \bibinfo {author} {\bibfnamefont {A.}~\bibnamefont {R{\'{o}}{\.{z}}anska}}, \bibinfo {author} {\bibfnamefont {T.~P.}\ \bibnamefont {Adhikari}}, \bibinfo {author} {\bibfnamefont {A.}~\bibnamefont {Pandey}}, \bibinfo {author} {\bibfnamefont {V.~K.}\ \bibnamefont {Jaiswal}}, \bibinfo {author} {\bibfnamefont {V.}~\bibnamefont {Karas}}, \bibinfo {author} {\bibfnamefont {A.}~\bibnamefont {Borkar}}, \bibinfo {author} {\bibfnamefont {M.~L.}\ \bibnamefont {Mart{\'{i}}nez-Aldama}}, \ and\ \bibinfo {author} {\bibfnamefont {R.}~\bibnamefont {Prince}},\ }\href {\doibase 10.1140/epjd/s10053-023-00630-8} {\bibfield  {journal} {\bibinfo  {journal} {Eur. Phys. J. D}\ }\textbf {\bibinfo
  {volume} {77}},\ \bibinfo {pages} {56} (\bibinfo {year} {2023})}\BibitemShut {NoStop}%
\bibitem [{\citenamefont {Sturrock}(1971)}]{Sturrock1971}%
  \BibitemOpen
  \bibfield  {author} {\bibinfo {author} {\bibfnamefont {P.~A.}\ \bibnamefont {Sturrock}},\ }\href@noop {} {\bibfield  {journal} {\bibinfo  {journal} {Astrophys. J.}\ }\textbf {\bibinfo {volume} {164}},\ \bibinfo {pages} {529} (\bibinfo {year} {1971})}\BibitemShut {NoStop}%
\bibitem [{\citenamefont {Ruderman}\ and\ \citenamefont {Sutherland}(1975)}]{Ruderman1975}%
  \BibitemOpen
  \bibfield  {author} {\bibinfo {author} {\bibfnamefont {M.~A.}\ \bibnamefont {Ruderman}}\ and\ \bibinfo {author} {\bibfnamefont {P.~G.}\ \bibnamefont {Sutherland}},\ }\href {\doibase 10.1086/153393} {\bibfield  {journal} {\bibinfo  {journal} {Astrophys. J.}\ }\textbf {\bibinfo {volume} {196}},\ \bibinfo {pages} {51} (\bibinfo {year} {1975})}\BibitemShut {NoStop}%
\bibitem [{\citenamefont {Michel}(1982)}]{Michel1982}%
  \BibitemOpen
  \bibfield  {author} {\bibinfo {author} {\bibfnamefont {F.~C.}\ \bibnamefont {Michel}},\ }\href {\doibase 10.1103/RevModPhys.54.1} {\bibfield  {journal} {\bibinfo  {journal} {Rev. Mod. Phys.}\ }\textbf {\bibinfo {volume} {54}},\ \bibinfo {pages} {1} (\bibinfo {year} {1982})}\BibitemShut {NoStop}%
\bibitem [{\citenamefont {Tajima}\ and\ \citenamefont {Shibata}(2019)}]{Tajima}%
  \BibitemOpen
  \bibfield  {author} {\bibinfo {author} {\bibfnamefont {T.}~\bibnamefont {Tajima}}\ and\ \bibinfo {author} {\bibfnamefont {K.}~\bibnamefont {Shibata}},\ }\href {https://books.google.com.pk/books?id=ORMOyAEACAAJ} {\emph {\bibinfo {title} {Plasma Astrophysics}}},\ Frontiers in Physics\ (\bibinfo  {publisher} {CRC Press Taylor \& Francis Group},\ \bibinfo {year} {2019})\BibitemShut {NoStop}%
\bibitem [{\citenamefont {Alfv{\'{e}}n}(1981)}]{Alfven1981}%
  \BibitemOpen
  \bibfield  {author} {\bibinfo {author} {\bibfnamefont {H.}~\bibnamefont {Alfv{\'{e}}n}},\ }\href {\doibase 10.1007/978-94-009-8374-8} {\emph {\bibinfo {title} {{Cosmic Plasma}}}},\ \bibinfo {edition} {1st}\ ed.\ (\bibinfo  {publisher} {Springer Netherlands},\ \bibinfo {address} {Dordrecht},\ \bibinfo {year} {1981})\BibitemShut {NoStop}%
\bibitem [{\citenamefont {Shukla}\ and\ \citenamefont {Marklund}(2004)}]{Shukla2004}%
  \BibitemOpen
  \bibfield  {author} {\bibinfo {author} {\bibfnamefont {P.~K.}\ \bibnamefont {Shukla}}\ and\ \bibinfo {author} {\bibfnamefont {M.}~\bibnamefont {Marklund}},\ }\href {\doibase 10.1238/Physica.Topical.113a00036} {\bibfield  {journal} {\bibinfo  {journal} {Phys. Scr.}\ }\textbf {\bibinfo {volume} {T113}},\ \bibinfo {pages} {36} (\bibinfo {year} {2004})}\BibitemShut {NoStop}%
\bibitem [{\citenamefont {Shukla}(2008)}]{Shukla2008}%
  \BibitemOpen
  \bibfield  {author} {\bibinfo {author} {\bibfnamefont {P.~K.}\ \bibnamefont {Shukla}},\ }\href {\doibase 10.1088/0031-8949/77/06/068201} {\bibfield  {journal} {\bibinfo  {journal} {Phys. Scr.}\ }\textbf {\bibinfo {volume} {77}},\ \bibinfo {pages} {068201} (\bibinfo {year} {2008})}\BibitemShut {NoStop}%
\bibitem [{\citenamefont {Higdon}, \citenamefont {Lingenfelter},\ and\ \citenamefont {Rothschild}(2009)}]{Higdon2009}%
  \BibitemOpen
  \bibfield  {author} {\bibinfo {author} {\bibfnamefont {J.~C.}\ \bibnamefont {Higdon}}, \bibinfo {author} {\bibfnamefont {R.~E.}\ \bibnamefont {Lingenfelter}}, \ and\ \bibinfo {author} {\bibfnamefont {R.~E.}\ \bibnamefont {Rothschild}},\ }\href {\doibase 10.1088/0004-637X/698/1/350} {\bibfield  {journal} {\bibinfo  {journal} {Astrophys. J.}\ }\textbf {\bibinfo {volume} {698}},\ \bibinfo {pages} {350} (\bibinfo {year} {2009})}\BibitemShut {NoStop}%
\bibitem [{\citenamefont {Gusev}\ \emph {et~al.}(2001)\citenamefont {Gusev}, \citenamefont {Jayanthi}, \citenamefont {Martin}, \citenamefont {Pugacheva},\ and\ \citenamefont {Spjeldvik}}]{Gusev2001}%
  \BibitemOpen
  \bibfield  {author} {\bibinfo {author} {\bibfnamefont {A.~A.}\ \bibnamefont {Gusev}}, \bibinfo {author} {\bibfnamefont {U.~B.}\ \bibnamefont {Jayanthi}}, \bibinfo {author} {\bibfnamefont {I.~M.}\ \bibnamefont {Martin}}, \bibinfo {author} {\bibfnamefont {G.~I.}\ \bibnamefont {Pugacheva}}, \ and\ \bibinfo {author} {\bibfnamefont {W.~N.}\ \bibnamefont {Spjeldvik}},\ }\href {\doibase 10.1029/1999JA000443} {\bibfield  {journal} {\bibinfo  {journal} {J. Geophys. Res. Sp. Phys.}\ }\textbf {\bibinfo {volume} {106}},\ \bibinfo {pages} {26111} (\bibinfo {year} {2001})}\BibitemShut {NoStop}%
\bibitem [{\citenamefont {Hor{\'{a}}nyi}(1996)}]{Horanyi1996}%
  \BibitemOpen
  \bibfield  {author} {\bibinfo {author} {\bibfnamefont {M.}~\bibnamefont {Hor{\'{a}}nyi}},\ }\href {\doibase 10.1146/annurev.astro.34.1.383} {\bibfield  {journal} {\bibinfo  {journal} {Annu. Rev. Astron. Astrophys.}\ }\textbf {\bibinfo {volume} {34}},\ \bibinfo {pages} {383} (\bibinfo {year} {1996})}\BibitemShut {NoStop}%
\bibitem [{\citenamefont {Surko}\ and\ \citenamefont {Murphy}(1990)}]{Surko1990}%
  \BibitemOpen
  \bibfield  {author} {\bibinfo {author} {\bibfnamefont {C.~M.}\ \bibnamefont {Surko}}\ and\ \bibinfo {author} {\bibfnamefont {T.~J.}\ \bibnamefont {Murphy}},\ }\href {\doibase 10.1063/1.859558} {\bibfield  {journal} {\bibinfo  {journal} {Phys. Fluids B Plasma Phys.}\ }\textbf {\bibinfo {volume} {2}},\ \bibinfo {pages} {1372} (\bibinfo {year} {1990})}\BibitemShut {NoStop}%
\bibitem [{\citenamefont {Krasheninnikov}\ and\ \citenamefont {Soboleva}(2005)}]{Krasheninnikov2005}%
  \BibitemOpen
  \bibfield  {author} {\bibinfo {author} {\bibfnamefont {S.~I.}\ \bibnamefont {Krasheninnikov}}\ and\ \bibinfo {author} {\bibfnamefont {T.~K.}\ \bibnamefont {Soboleva}},\ }\href {\doibase 10.1088/0741-3335/47/5A/025} {\bibfield  {journal} {\bibinfo  {journal} {Plasma Phys. Control. Fusion}\ }\textbf {\bibinfo {volume} {47}},\ \bibinfo {pages} {A339} (\bibinfo {year} {2005})}\BibitemShut {NoStop}%
\bibitem [{\citenamefont {Guanying}\ \emph {et~al.}(2017)\citenamefont {Guanying}, \citenamefont {Liu}, \citenamefont {Xie},\ and\ \citenamefont {Li}}]{Guanying2017}%
  \BibitemOpen
  \bibfield  {author} {\bibinfo {author} {\bibfnamefont {Y.}~\bibnamefont {Guanying}}, \bibinfo {author} {\bibfnamefont {J.}~\bibnamefont {Liu}}, \bibinfo {author} {\bibfnamefont {J.}~\bibnamefont {Xie}}, \ and\ \bibinfo {author} {\bibfnamefont {J.}~\bibnamefont {Li}},\ }\href {\doibase 10.1016/j.fusengdes.2017.03.144} {\bibfield  {journal} {\bibinfo  {journal} {Fusion Eng. Des.}\ }\textbf {\bibinfo {volume} {118}},\ \bibinfo {pages} {124} (\bibinfo {year} {2017})}\BibitemShut {NoStop}%
\bibitem [{\citenamefont {Banerjee}\ and\ \citenamefont {Maitra}(2016)}]{Banerjee2016}%
  \BibitemOpen
  \bibfield  {author} {\bibinfo {author} {\bibfnamefont {G.}~\bibnamefont {Banerjee}}\ and\ \bibinfo {author} {\bibfnamefont {S.}~\bibnamefont {Maitra}},\ }\href {\doibase 10.1063/1.4971223} {\bibfield  {journal} {\bibinfo  {journal} {Phys. Plasmas}\ }\textbf {\bibinfo {volume} {23}},\ \bibinfo {pages} {123701} (\bibinfo {year} {2016})}\BibitemShut {NoStop}%
\bibitem [{\citenamefont {Paul}, \citenamefont {Bandyopadhyay},\ and\ \citenamefont {Das}(2017)}]{Paul2017}%
  \BibitemOpen
  \bibfield  {author} {\bibinfo {author} {\bibfnamefont {A.}~\bibnamefont {Paul}}, \bibinfo {author} {\bibfnamefont {A.}~\bibnamefont {Bandyopadhyay}}, \ and\ \bibinfo {author} {\bibfnamefont {K.~P.}\ \bibnamefont {Das}},\ }\href {\doibase 10.1063/1.4975089} {\bibfield  {journal} {\bibinfo  {journal} {Phys. Plasmas}\ }\textbf {\bibinfo {volume} {24}},\ \bibinfo {pages} {013707} (\bibinfo {year} {2017})}\BibitemShut {NoStop}%
\bibitem [{\citenamefont {Singh}, \citenamefont {Kaur},\ and\ \citenamefont {Saini}(2017)}]{Singh2017}%
  \BibitemOpen
  \bibfield  {author} {\bibinfo {author} {\bibfnamefont {K.}~\bibnamefont {Singh}}, \bibinfo {author} {\bibfnamefont {N.}~\bibnamefont {Kaur}}, \ and\ \bibinfo {author} {\bibfnamefont {N.~S.}\ \bibnamefont {Saini}},\ }\href {\doibase 10.1063/1.4984996} {\bibfield  {journal} {\bibinfo  {journal} {Phys. Plasmas}\ }\textbf {\bibinfo {volume} {24}},\ \bibinfo {pages} {063703} (\bibinfo {year} {2017})}\BibitemShut {NoStop}%
\bibitem [{\citenamefont {El-Kalaawy}(2018)}]{El-kalaawy2018}%
  \BibitemOpen
  \bibfield  {author} {\bibinfo {author} {\bibfnamefont {O.~H.}\ \bibnamefont {El-Kalaawy}},\ }\href {\doibase 10.1140/epjp/i2018-11873-7} {\bibfield  {journal} {\bibinfo  {journal} {Eur. Phys. J. Plus}\ }\textbf {\bibinfo {volume} {133}},\ \bibinfo {pages} {58} (\bibinfo {year} {2018})}\BibitemShut {NoStop}%
\bibitem [{\citenamefont {Dev}\ \emph {et~al.}(2020)\citenamefont {Dev}, \citenamefont {Deka}, \citenamefont {Kalita},\ and\ \citenamefont {Sarma}}]{Dev2020}%
  \BibitemOpen
  \bibfield  {author} {\bibinfo {author} {\bibfnamefont {A.~N.}\ \bibnamefont {Dev}}, \bibinfo {author} {\bibfnamefont {M.~K.}\ \bibnamefont {Deka}}, \bibinfo {author} {\bibfnamefont {R.~K.}\ \bibnamefont {Kalita}}, \ and\ \bibinfo {author} {\bibfnamefont {J.}~\bibnamefont {Sarma}},\ }\href {\doibase 10.1140/epjp/s13360-020-00861-3} {\bibfield  {journal} {\bibinfo  {journal} {Eur. Phys. J. Plus}\ }\textbf {\bibinfo {volume} {135}},\ \bibinfo {pages} {843} (\bibinfo {year} {2020})}\BibitemShut {NoStop}%
\bibitem [{\citenamefont {Haque}\ and\ \citenamefont {Mannan}(2020)}]{Haque2020}%
  \BibitemOpen
  \bibfield  {author} {\bibinfo {author} {\bibfnamefont {M.~N.}\ \bibnamefont {Haque}}\ and\ \bibinfo {author} {\bibfnamefont {A.}~\bibnamefont {Mannan}},\ }\href {\doibase 10.1109/TPS.2020.3002424} {\bibfield  {journal} {\bibinfo  {journal} {IEEE Trans. Plasma Sci.}\ }\textbf {\bibinfo {volume} {48}},\ \bibinfo {pages} {2591} (\bibinfo {year} {2020})}\BibitemShut {NoStop}%
\bibitem [{\citenamefont {Rahman}\ \emph {et~al.}(2021)\citenamefont {Rahman}, \citenamefont {Chowdhury}, \citenamefont {Mannan},\ and\ \citenamefont {Mamun}}]{Rahman2021}%
  \BibitemOpen
  \bibfield  {author} {\bibinfo {author} {\bibfnamefont {M.~H.}\ \bibnamefont {Rahman}}, \bibinfo {author} {\bibfnamefont {N.~A.}\ \bibnamefont {Chowdhury}}, \bibinfo {author} {\bibfnamefont {A.}~\bibnamefont {Mannan}}, \ and\ \bibinfo {author} {\bibfnamefont {A.~A.}\ \bibnamefont {Mamun}},\ }\href {\doibase 10.3390/galaxies9020031} {\bibfield  {journal} {\bibinfo  {journal} {Galaxies}\ }\textbf {\bibinfo {volume} {9}},\ \bibinfo {pages} {31} (\bibinfo {year} {2021})}\BibitemShut {NoStop}%
\bibitem [{\citenamefont {Halder}\ \emph {et~al.}(2023)\citenamefont {Halder}, \citenamefont {Dalui}, \citenamefont {Sardar},\ and\ \citenamefont {Bandyopadhyay}}]{Halder2023}%
  \BibitemOpen
  \bibfield  {author} {\bibinfo {author} {\bibfnamefont {P.}~\bibnamefont {Halder}}, \bibinfo {author} {\bibfnamefont {S.}~\bibnamefont {Dalui}}, \bibinfo {author} {\bibfnamefont {S.}~\bibnamefont {Sardar}}, \ and\ \bibinfo {author} {\bibfnamefont {A.}~\bibnamefont {Bandyopadhyay}},\ }\href {\doibase 10.1140/epjp/s13360-023-04359-6} {\bibfield  {journal} {\bibinfo  {journal} {Eur. Phys. J. Plus}\ }\textbf {\bibinfo {volume} {138}},\ \bibinfo {pages} {734} (\bibinfo {year} {2023})}\BibitemShut {NoStop}%
\bibitem [{\citenamefont {Krause}\ and\ \citenamefont {R{\"{a}}dler}(1980)}]{Krause1980}%
  \BibitemOpen
  \bibfield  {author} {\bibinfo {author} {\bibfnamefont {F.}~\bibnamefont {Krause}}\ and\ \bibinfo {author} {\bibfnamefont {K.-H.}\ \bibnamefont {R{\"{a}}dler}},\ }\href {\doibase doi:10.1515/9783112729694} {\emph {\bibinfo {title} {{Mean-Field Magnetohydrodynamics and Dynamo Theory}}}}\ (\bibinfo  {publisher} {De Gruyter},\ \bibinfo {address} {Berlin, Boston},\ \bibinfo {year} {1980})\BibitemShut {NoStop}%
\bibitem [{\citenamefont {Mahajan}\ and\ \citenamefont {Yoshida}(1998)}]{Mahajan1998}%
  \BibitemOpen
  \bibfield  {author} {\bibinfo {author} {\bibfnamefont {S.~M.}\ \bibnamefont {Mahajan}}\ and\ \bibinfo {author} {\bibfnamefont {Z.}~\bibnamefont {Yoshida}},\ }\href {\doibase 10.1103/PhysRevLett.81.4863} {\bibfield  {journal} {\bibinfo  {journal} {Phys. Rev. Lett.}\ }\textbf {\bibinfo {volume} {81}},\ \bibinfo {pages} {4863} (\bibinfo {year} {1998})}\BibitemShut {NoStop}%
\bibitem [{\citenamefont {Ohsaki}\ \emph {et~al.}(2002)\citenamefont {Ohsaki}, \citenamefont {Shatashvili}, \citenamefont {Yoshida},\ and\ \citenamefont {Mahajan}}]{Ohsaki2002}%
  \BibitemOpen
  \bibfield  {author} {\bibinfo {author} {\bibfnamefont {S.}~\bibnamefont {Ohsaki}}, \bibinfo {author} {\bibfnamefont {N.~L.}\ \bibnamefont {Shatashvili}}, \bibinfo {author} {\bibfnamefont {Z.}~\bibnamefont {Yoshida}}, \ and\ \bibinfo {author} {\bibfnamefont {S.~M.}\ \bibnamefont {Mahajan}},\ }\href {\doibase 10.1086/339499} {\bibfield  {journal} {\bibinfo  {journal} {Astrophys. J.}\ }\textbf {\bibinfo {volume} {570}},\ \bibinfo {pages} {395} (\bibinfo {year} {2002})}\BibitemShut {NoStop}%
\bibitem [{\citenamefont {Krishan}(1997)}]{Krishan1997}%
  \BibitemOpen
  \bibfield  {author} {\bibinfo {author} {\bibfnamefont {V.}~\bibnamefont {Krishan}},\ }\href {\doibase 10.1023/a:1004922327108} {\bibfield  {journal} {\bibinfo  {journal} {Space Sci. Rev.}\ }\textbf {\bibinfo {volume} {80}},\ \bibinfo {pages} {445} (\bibinfo {year} {1997})}\BibitemShut {NoStop}%
\bibitem [{\citenamefont {Asano}\ and\ \citenamefont {Takahara}(2007)}]{Asano2007}%
  \BibitemOpen
  \bibfield  {author} {\bibinfo {author} {\bibfnamefont {K.}~\bibnamefont {Asano}}\ and\ \bibinfo {author} {\bibfnamefont {F.}~\bibnamefont {Takahara}},\ }\href {\doibase 10.1086/509756} {\bibfield  {journal} {\bibinfo  {journal} {Astrophys. J.}\ }\textbf {\bibinfo {volume} {655}},\ \bibinfo {pages} {762} (\bibinfo {year} {2007})}\BibitemShut {NoStop}%
\bibitem [{\citenamefont {Foschini}(2002)}]{Foschini2002}%
  \BibitemOpen
  \bibfield  {author} {\bibinfo {author} {\bibfnamefont {L.}~\bibnamefont {Foschini}},\ }\href {\doibase 10.1051/0004-6361:20020231} {\bibfield  {journal} {\bibinfo  {journal} {Astron. Astrophys.}\ }\textbf {\bibinfo {volume} {385}},\ \bibinfo {pages} {62} (\bibinfo {year} {2002})}\BibitemShut {NoStop}%
\bibitem [{\citenamefont {Chakrabarti}, \citenamefont {Rosner},\ and\ \citenamefont {Vainshtein}(1994)}]{Chakrabarti1994}%
  \BibitemOpen
  \bibfield  {author} {\bibinfo {author} {\bibfnamefont {S.~K.}\ \bibnamefont {Chakrabarti}}, \bibinfo {author} {\bibfnamefont {R.}~\bibnamefont {Rosner}}, \ and\ \bibinfo {author} {\bibfnamefont {S.~I.}\ \bibnamefont {Vainshtein}},\ }\href {\doibase 10.1038/368434a0} {\bibfield  {journal} {\bibinfo  {journal} {Nature}\ }\textbf {\bibinfo {volume} {368}},\ \bibinfo {pages} {434} (\bibinfo {year} {1994})}\BibitemShut {NoStop}%
\bibitem [{\citenamefont {{Colgate}}\ and\ \citenamefont {{Li}}(1997)}]{Colgate1997}%
  \BibitemOpen
  \bibfield  {author} {\bibinfo {author} {\bibfnamefont {S.~A.}\ \bibnamefont {{Colgate}}}\ and\ \bibinfo {author} {\bibfnamefont {H.}~\bibnamefont {{Li}}},\ }in\ \href@noop {} {\emph {\bibinfo {booktitle} {Relativistic Jets in AGNs}}},\ \bibinfo {editor} {edited by\ \bibinfo {editor} {\bibfnamefont {M.}~\bibnamefont {{Ostrowski}}}, \bibinfo {editor} {\bibfnamefont {M.}~\bibnamefont {{Sikora}}}, \bibinfo {editor} {\bibfnamefont {G.}~\bibnamefont {{Madejski}}}, \ and\ \bibinfo {editor} {\bibfnamefont {M.}~\bibnamefont {{Begelman}}}}\ (\bibinfo {year} {1997})\ pp.\ \bibinfo {pages} {170--179}\BibitemShut {NoStop}%
\bibitem [{\citenamefont {Bao}\ \emph {et~al.}(1997)\citenamefont {Bao}, \citenamefont {Hadrava}, \citenamefont {Wiita},\ and\ \citenamefont {Xiong}}]{Bao1997}%
  \BibitemOpen
  \bibfield  {author} {\bibinfo {author} {\bibfnamefont {G.}~\bibnamefont {Bao}}, \bibinfo {author} {\bibfnamefont {P.}~\bibnamefont {Hadrava}}, \bibinfo {author} {\bibfnamefont {P.~J.}\ \bibnamefont {Wiita}}, \ and\ \bibinfo {author} {\bibfnamefont {Y.}~\bibnamefont {Xiong}},\ }\href {\doibase 10.1086/304575} {\bibfield  {journal} {\bibinfo  {journal} {Astrophys. J.}\ }\textbf {\bibinfo {volume} {487}},\ \bibinfo {pages} {142} (\bibinfo {year} {1997})}\BibitemShut {NoStop}%
\bibitem [{\citenamefont {Pariev}\ and\ \citenamefont {Colgate}(2007)}]{Pariev2007a}%
  \BibitemOpen
  \bibfield  {author} {\bibinfo {author} {\bibfnamefont {V.~I.}\ \bibnamefont {Pariev}}\ and\ \bibinfo {author} {\bibfnamefont {S.~A.}\ \bibnamefont {Colgate}},\ }\href {\doibase 10.1086/510734} {\bibfield  {journal} {\bibinfo  {journal} {Astrophys. J.}\ }\textbf {\bibinfo {volume} {658}},\ \bibinfo {pages} {114} (\bibinfo {year} {2007})}\BibitemShut {NoStop}%
\bibitem [{\citenamefont {Pariev}, \citenamefont {Colgate},\ and\ \citenamefont {Finn}(2007)}]{Pariev2007b}%
  \BibitemOpen
  \bibfield  {author} {\bibinfo {author} {\bibfnamefont {V.~I.}\ \bibnamefont {Pariev}}, \bibinfo {author} {\bibfnamefont {S.~A.}\ \bibnamefont {Colgate}}, \ and\ \bibinfo {author} {\bibfnamefont {J.~M.}\ \bibnamefont {Finn}},\ }\href {\doibase 10.1086/510735} {\bibfield  {journal} {\bibinfo  {journal} {Astrophys. J.}\ }\textbf {\bibinfo {volume} {658}},\ \bibinfo {pages} {129} (\bibinfo {year} {2007})}\BibitemShut {NoStop}%
\bibitem [{\citenamefont {Crenshaw}, \citenamefont {Kraemer},\ and\ \citenamefont {George}(2003)}]{Crenshaw2003}%
  \BibitemOpen
  \bibfield  {author} {\bibinfo {author} {\bibfnamefont {D.~M.}\ \bibnamefont {Crenshaw}}, \bibinfo {author} {\bibfnamefont {S.~B.}\ \bibnamefont {Kraemer}}, \ and\ \bibinfo {author} {\bibfnamefont {I.~M.}\ \bibnamefont {George}},\ }\href {\doibase 10.1146/annurev.astro.41.082801.100328} {\bibfield  {journal} {\bibinfo  {journal} {Annu. Rev. Astron. Astrophys.}\ }\textbf {\bibinfo {volume} {41}},\ \bibinfo {pages} {117} (\bibinfo {year} {2003})}\BibitemShut {NoStop}%
\bibitem [{\citenamefont {Blandford}\ and\ \citenamefont {Payne}(1982)}]{Blandford1982}%
  \BibitemOpen
  \bibfield  {author} {\bibinfo {author} {\bibfnamefont {R.~D.}\ \bibnamefont {Blandford}}\ and\ \bibinfo {author} {\bibfnamefont {D.~G.}\ \bibnamefont {Payne}},\ }\href {\doibase 10.1093/mnras/199.4.883} {\bibfield  {journal} {\bibinfo  {journal} {Mon. Not. R. Astron. Soc.}\ }\textbf {\bibinfo {volume} {199}},\ \bibinfo {pages} {883} (\bibinfo {year} {1982})}\BibitemShut {NoStop}%
\bibitem [{\citenamefont {Contopoulos}\ and\ \citenamefont {Lovelace}(1994)}]{Contopoulos1994}%
  \BibitemOpen
  \bibfield  {author} {\bibinfo {author} {\bibfnamefont {J.}~\bibnamefont {Contopoulos}}\ and\ \bibinfo {author} {\bibfnamefont {R.~V.~E.}\ \bibnamefont {Lovelace}},\ }\href {\doibase 10.1086/174307} {\bibfield  {journal} {\bibinfo  {journal} {Astrophys. J.}\ }\textbf {\bibinfo {volume} {429}},\ \bibinfo {pages} {139} (\bibinfo {year} {1994})}\BibitemShut {NoStop}%
\bibitem [{\citenamefont {Lopez-Rodriguez}\ \emph {et~al.}(2018)\citenamefont {Lopez-Rodriguez}, \citenamefont {Alonso-Herrero}, \citenamefont {Diaz-Santos}, \citenamefont {Gonzalez-Martin}, \citenamefont {Ichikawa}, \citenamefont {Levenson}, \citenamefont {Martinez-Paredes}, \citenamefont {Nikutta}, \citenamefont {Packham}, \citenamefont {Perlman}, \citenamefont {{Ramos Almeida}}, \citenamefont {Rodriguez-Espinosa},\ and\ \citenamefont {Telesco}}]{Lopez2018}%
  \BibitemOpen
  \bibfield  {author} {\bibinfo {author} {\bibfnamefont {E.}~\bibnamefont {Lopez-Rodriguez}}, \bibinfo {author} {\bibfnamefont {A.}~\bibnamefont {Alonso-Herrero}}, \bibinfo {author} {\bibfnamefont {T.}~\bibnamefont {Diaz-Santos}}, \bibinfo {author} {\bibfnamefont {O.}~\bibnamefont {Gonzalez-Martin}}, \bibinfo {author} {\bibfnamefont {K.}~\bibnamefont {Ichikawa}}, \bibinfo {author} {\bibfnamefont {N.~A.}\ \bibnamefont {Levenson}}, \bibinfo {author} {\bibfnamefont {M.}~\bibnamefont {Martinez-Paredes}}, \bibinfo {author} {\bibfnamefont {R.}~\bibnamefont {Nikutta}}, \bibinfo {author} {\bibfnamefont {C.}~\bibnamefont {Packham}}, \bibinfo {author} {\bibfnamefont {E.}~\bibnamefont {Perlman}}, \bibinfo {author} {\bibfnamefont {C.}~\bibnamefont {{Ramos Almeida}}}, \bibinfo {author} {\bibfnamefont {J.~M.}\ \bibnamefont {Rodriguez-Espinosa}}, \ and\ \bibinfo {author} {\bibfnamefont {C.~M.}\ \bibnamefont {Telesco}},\ }\href {\doibase 10.1093/mnras/sty1197} {\bibfield  {journal} {\bibinfo  {journal} {Mon. Not. R. Astron.
  Soc.}\ }\textbf {\bibinfo {volume} {478}},\ \bibinfo {pages} {2350} (\bibinfo {year} {2018})}\BibitemShut {NoStop}%
\bibitem [{\citenamefont {Vlahakis}\ and\ \citenamefont {Konigl}(2004)}]{Vlahakis2004}%
  \BibitemOpen
  \bibfield  {author} {\bibinfo {author} {\bibfnamefont {N.}~\bibnamefont {Vlahakis}}\ and\ \bibinfo {author} {\bibfnamefont {A.}~\bibnamefont {Konigl}},\ }\href {\doibase 10.1086/382670} {\bibfield  {journal} {\bibinfo  {journal} {Astrophys. J.}\ }\textbf {\bibinfo {volume} {605}},\ \bibinfo {pages} {656} (\bibinfo {year} {2004})}\BibitemShut {NoStop}%
\end{thebibliography}%
\end{document}